\documentclass[twocolumn]{aastex631}

\graphicspath{{./}{figures/}}
\usepackage{amsmath}

\usepackage{tensor}

\usepackage{CJK}

\usepackage[T1]{fontenc}

\newcommand{\R}{\mathcal{R}}

\shorttitle{Rotational and Magnetic Effects on BH Accretion Variability}
\shortauthors{Cho \& Narayan}

\begin{document}
\begin{CJK}{UTF8}{mj}
\title{Variability in Black Hole Accretion: Dependence on Rotational and Magnetic Energy Balance}

\author[0000-0002-2858-9481]{Hyerin Cho (조혜린)}
\affiliation{Center for Astrophysics $\vert$ Harvard \& Smithsonian, 60 Garden Street, Cambridge, MA 02138, USA}
\affiliation{Black Hole Initiative at Harvard University, 20 Garden Street, Cambridge, MA 02138, USA}

\author[0000-0002-1919-2730]{Ramesh Narayan}
\affiliation{Center for Astrophysics $\vert$ Harvard \& Smithsonian, 60 Garden Street, Cambridge, MA 02138, USA}
\affiliation{Black Hole Initiative at Harvard University, 20 Garden Street, Cambridge, MA 02138, USA}

\begin{abstract} \label{abstract}
Most general relativistic magnetohydrodynamic simulations of black hole (BH) hot accretion flows are initialized with small rotating tori and produce stable jets with only small fluctuations. However, recent studies using larger scale Bondi-like initial conditions have reported intermittent jet activity and loss of coherent rotation.  To investigate the differences, we modify the standard torus setup across four BH spins: $a_*=0$, $0.5$, $0.9$, $-0.9$. First, we increase the torus size significantly (pressure maximum at 500 gravitational radii), allowing long simulations ($2.8\times10^5$ gravitational times) without gas depletion. These runs reproduce the weak variability seen in smaller tori, indicating that a larger dynamic range alone does not cause strong fluctuations. We observe moderate suppression of the accretion rate
by factors of $\sim 1.6, ~2.5$ for BH spins $a_*=0.5,~0.9$, respectively, compared to $a_*=0$. Also, the density profile scales as $\rho(r)\propto r^{-1.1}$ for prograde BHs. Next, we considerably strengthen the initial magnetic field in the large torus by setting the plasma-$\beta\approx 1$. This induces strong variability in the evolution. The jet efficiency in the $a_*=0.9$ model, for instance, now varies by over 3 orders of magnitude, and gas rotation reverses directions. Combining these results with prior studies, we propose that a key parameter is the ratio $\R$ between the rotational and magnetic energies in the initial state. Strong variability appears later in models with a larger value of $\R$. The implication is that all simulations, and by extension all hot accretion flows in Nature, will ultimately develop intermittent jets if evolved long enough.

\end{abstract}

\section{Introduction} \label{sec:intro}
Black holes (BHs) accreting at highly sub-Eddington rates invariably have hot geometrically thick accretion flows (\citealt{Narayan1994,Narayan1995,Abramowicz1995}, see also \citealt{Shapiro1976,Ichimaru1977,Rees1982} and \citealt{Yuan2014} for a review). These systems are radiatively inefficient (hence they are sometimes referred to as radiatively inefficient accretion flows, RIAFs), or equivalently, they advect most of their internal energy (hence they are also known as advection-dominated accretion flows, ADAFs). 

While early models of hot accretion flows were purely hydrodynamical, the importance of magnetic fields was soon appreciated, and this led to the recognition of two  magnetic sub-states: standard and normal evolution (SANE) systems with weaker fields, and magnetically arrested disk (MAD) systems with strong fields. The MAD state (\citealt{Igumenshchev2003,Narayan2003,Tchekhovskoy2011}, see also \citealt{Shvartsman1971,Bisnovatyi-Kogan1974,Bisnovatyi1976}) is of particular interest, as it seems that many, if not most, hot accretion flows around supermassive BHs might occur in this regime \citep[e.g.,][]{Narayan2003,Narayan2022}. In MAD accretion, frozen-in magnetic fields in the accreting plasma accumulate at the center and the field strength builds up until the field becomes dynamically important and hinders gas accretion \citep{Narayan2003}. Axisymmetry is then inevitably broken via Rayleigh-Taylor (or interchange) instability, and the gas accretes via narrow streams that squeeze through gaps between strong magnetic field bundles. 

Hot accretion flows around spinning black holes produce relativistic jets \citep{McKinney2004,Tchekhovskoy2011}, which appear to be launched via the \citet{Blandford1977} (BZ) mechanism. The jet is generated when the spinning BH spacetime frame-drags polodial magnetic field lines in the vicinity of the horizon (especially in the ergo-region), thereby enabling the jet to tap into the BH's spin energy. The energy outflow in the jet can be powerful enough to impact conditions in the surrounding host galaxy and even beyond, a process that is generally referred to as BH feedback \citep[e.g.,][]{McNamara2007,McNamara2012, Fabian2012, Heckman2014,Blandford2019}.

General relativistic magneto-hydrodynamic (GRMHD) simulations are commonly used to study hot accretion flows. Because these flows are radiatively inefficient, one can neglect radiative cooling altogether, which makes simulations scale-free, i.e., one can run a single simulation for an arbitrary BH mass $M$ and accretion rate $\dot{M}$ and then simply scale the results for other masses and accretion rates. The same is not true for the dimensionless spin parameter $a_* \equiv a/M$ of the BH; each value of $a_*$ needs to be simulated individually. 

GRMHD simulations have demonstrated that the MAD state emerges naturally, and a BZ-powered jet is launched self-consistently, from fairly generic initial conditions (ICs)  \citep[][to name a few]{McKinney2004, Tchekhovskoy2011, Lasota2014, Narayan2022}. In an important study, \citet{Tchekhovskoy2011} reported a GRMHD simulation of a MAD flow around a rapidly spinning BH, where the jet power $P_J$ was greater than the entire accreted rest-mass energy: $P_J > \dot{M}c^2$. This was the first incontrovertible, simulation-based evidence that the jets extract at least some of their power directly from the BH's spin energy. 

GRMHD simulations have played an essential role in interpreting the first BH images from the Event Horizon Telescope (EHT). A comparison of simulation data with EHT observations suggests that, currently, accretion in both M87* \citep{EHTC2021} and Sgr A* \citep{EHTC2022} is likely operating in the MAD regime.

Solving the general relativistic equations of magnetohydrodynamics with high spatial resolution on BH event horizon scales is computationally expensive. Because of this, GRMHD simulations have typically focused on a single class of IC, a \citet[][hereafter FM]{Fishbone1976} torus, and have covered only a limited range of scales  \citep[see][for reviews]{Davis2020,Komissarov2021}; typically, inflow equilibrium (steady state) extends from the event horizon $r_H$ out to only a few tens of $r_H$. The rotating FM torus solution in the simulations is initially in hydrodynamic equilibrium and embedded with a weak poloidal magnetic field. As the simulation evolves, the field is amplified by the magnetorotational instability (MRI) and the strong field then takes charge of the gas dynamics. The characteristic length-scale of the initial torus is often chosen to be small, e.g., the radius of the pressure maximum in the torus is $r_{\rm max} \sim 10-20 \,r_g$ \citep[e.g.,][]{Porth:2019}. Such a small torus size only allows a limited budget of gas and magnetic field for accretion. Over the course of the simulation, gas and magnetic fields drift out of the simulation domain into the BH (some also escapes at large radii), which eventually starves the BH. In contrast, many BHs in the real universe are embedded in rather different environments filled with an essentially infinite supply of gas and magnetic fields.

Recent works have started exploring ICs which are different from the FM torus and closer to real systems in nature \citep{Ressler2020, Ressler2021, Lalakos2022, Kwan2023, Kaaz2023, Cho2023, Cho2024, Kaaz2025, Galishnikova2025, Kim2025, Guo2025, Lalakos2025}. The diverse ICs that have been explored include modeling stellar winds near Sgr A*, \citet{Bondi1952}-type spherically symmetric accretion, re-mapping a realistic environment from galaxy simulations, initializing magnetic fields with various strengths and geometries, and experimenting with different levels of gas rotation. For the purposes of this paper, we will refer to this whole class of simulations as `Bondi-like' setups to distinguish them from the more idealized rapidly rotating `torus' simulations that have dominated much of the GRMHD literature. As in the case of torus simulations, it is found that the MAD state is reached generically even in Bondi-like simulations. However, some notable differences have been identified. Understanding the differences is one of the motivations of the present paper. 

The most distinctive result to date is that some simulations with spinning BHs and Bondi-like setups exhibit intermittent jet activity \citep{Ressler2021,Kwan2023,Lalakos2024, Galishnikova2025,Kim2025,Guo2025,Lalakos2025}\footnote{\citet{Chan2025} also found similar results, but with modified FM torus ICs.}. Another interesting result is the loss of coherent rotation in the accreting gas \citep{Cho2024}. These features have not been observed in idealized torus simulations.
Due to the many differences between Bondi-like and torus simulations, e.g., spanning a large vs small dynamic range of radius, using strong vs weak initial magnetic fields, initializing with weakly/non-rotating gas vs strongly rotating gas, etc. (summarized in \autoref{tab:comparison_prev_work}, and discussed in detail in Section~\ref{sec:comparison}), it is challenging to disentangle which particular difference in initial setup  causes which deviation in the final results. Nevertheless, some patterns are emerging. For instance, Bondi-like
simulations which consider a wide range of radii and begin with non- or weakly rotating gas generally find intermittent jet activity \citep{Ressler2021,Kwan2023,Lalakos2024,Galishnikova2025,Kim2025,Guo2025,Chan2025,Lalakos2025}. 
Also, in simulations initialized with strong magnetic fields ($\beta\sim 1$, the plasma-$\beta$ is defined in \autoref{eq:plasma beta}) such that the magnetic field is dynamically important right from the beginning, coherent rotation is disrupted \citep{Cho2024}, likely due to efficient removal of angular momentum via magnetic flux eruptions \citep{Chatterjee2022}. However, a similar effect has never been seen in a conventional small FM torus simulation.

As a first attempt to understand the above differences between torus and Bondi-like simulations, the present paper presents torus simulations in which we modify the initial FM torus in ways to emulate more closely the Bondi-like simulations. After describing the numerical methods in Section~\ref{sec:numerical_method}, we consider in Section~\ref{sec:standard models} initial tori which are more than a factor of 10 larger than the tori used in most previous GRMHD simulations. By thus expanding the dynamic range of spatial scales, we explore whether a larger dynamic range by itself can explain some of the surprising new results from the Bondi-like runs. Then, in Section~\ref{sec:beta1_models}, we increase the strength of the magnetic field in the initial torus to explore what effect this might have on the subsequent steady state accretion flow. In Section~\ref{sec:comparison}, we compare our modified FM torus simulations in the context of other Bondi-like simulations and search for common trends. We summarize and conclude in Section~\ref{sec:conclude}.

Throughout, we adopt units in which $G=c=M_\bullet=1$, where $M_\bullet$ is the BH mass. Correspondingly, we measure lengths and times in units of the gravitational radius, $r_g\equiv GM_\bullet /c^2$, and gravitational time, $t_g\equiv r_g/c$.

\section{Numerical Methods}\label{sec:numerical_method}

The simulations reported here are run with the GRMHD code KORAL \citep{Sadowski2013,Sadowski2014}, which follows the methods described in \citet{Komissarov1999} and \citet{Gammie2003}. The code solves the ideal GRMHD equations of mass conservation and energy-momentum conservation,
\begin{equation}
(\rho u^\mu)_{;\mu}=0,
\end{equation}
\begin{equation} T^{\mu\nu}_{\ \ ;\mu}=0,
\end{equation}
and the ideal MHD induction equation, 
\begin{equation}
F^{*\mu\nu}_{\ \ ;\nu}=0,
\end{equation}
where $\rho$ is the rest mass density of the gas, $u^\mu$ is its 4-velocity, $T^{\mu\nu}$ is the stress-energy tensor,
\begin{equation}
T^{\mu\nu}=(\rho+u+p_g+b^2)u^\mu u^\nu + (p_g+b^2/2) g^{\mu\nu} -b^{\mu}b^{\nu},
\end{equation}
and $F^{*\mu\nu}$ is the dual of the electromagnetic field tensor. The magnetic field 4-vector $b^\mu$ is defined as \citep{Komissarov1999}
\begin{equation}
b^\mu \equiv \frac{1}{2} \epsilon^{\mu\nu\kappa\lambda} u_\nu F_{\lambda\kappa},
\end{equation}
and $b^2 = b^\mu b_\nu$.
The internal energy density $u$  and pressure $p_g$ of the gas are taken to be related by
\begin{equation}
p_g=(\gamma_{\rm ad}-1)u,
\end{equation}
with the adiabatic index set to $\gamma_{\rm ad} = 13/9$.\footnote{This is a standard choice in this field, though \citet{Gammie2025} has recently argued that $\gamma_{\rm ad} = 5/3$ might be a better choice for hot two-temperature accretion flows around BHs.} The gas temperature is $T\equiv p_g/\rho$, and the magnetic pressure is $p_b\equiv b^2/2$. The plasma-$\beta$ parameter is defined as the gas-to-magnetic pressure ratio, 
\begin{equation}
\beta\equiv \frac{p_g}{p_b}.
\label{eq:plasma beta}
\end{equation}
The simulations are run in a time-independent spinning spacetime, described by the Kerr metric with dimensionless BH spin parameter $a_*$. The (outer) event horizon is located at
\begin{equation}
r_H \equiv (1+\sqrt{1-a_*^2})\,r_g,
\end{equation}
and we use horizon-penetrating Kerr-Schild coordinates.

All the simulations described in this paper are initialized with a FM torus using an identical setup as in \citet{Narayan2022}, except that the torus in the present runs is much larger. The inner edge and pressure maximum of the torus are set to $r_{\rm in} = 255\,r_g$, $r_{\rm max} = 500\,r_g$, compared to $r_{\rm in}=20\,r_g$ and $r_{\rm max} \sim42-43\,r_g$ in \citet{Narayan2022} or the even smaller $r_{\rm in}=6\,r_g$ and $r_{\rm max} \sim12\,r_g$ in \citet{Porth:2019}.  Correspondingly, the initial loop of poloidal magnetic field has a characteristic radius $r_{\rm mag} = 2000\,r_g$, instead of $400\,r_g$ in \citet{Narayan2022}. The torus has its outer edge at $\approx 10^4\,r_g$. 

The initial magnetic field strength is parameterized by \citep[e.g.,][]{Porth:2019}
\begin{equation}
\beta_{\rm max}\equiv \frac{(p_g)_{\rm max}}{(p_b)_{\rm max}}~,
\end{equation}
where $(p_g)_{\rm max}$ and $(p_b)_{\rm max}$ are the
maximum gas pressure and magnetic pressure in the initial torus. In the runs discussed in Section~\ref{sec:standard models}, the initial magnetic field is weak, with $\beta_{\rm max}=100$, same as in \citet{Narayan2022}. In  the runs discussed in Section~\ref{sec:beta1_models}, however, the initial magnetic field is much stronger, $\beta_{\rm max}=1$. 

We employ the coordinate system and grid of \citet{Ressler2017}, with the same set of parameters as in \citet{Narayan2022}. The grid resolution in $(r,\theta,\varphi)$ for the simulations described in Section~\ref{sec:standard models} is $288\times 192\times 144$, and for those in Section~\ref{sec:beta1_models} is $192\times 128\times 128$. 

We consider four  choices of the BH spin parameter: $a_*=0, ~0.5, ~0.9, ~-0.9$. All simulations are run for a duration of $t=2.8\times 10^5\,t_g$. Because of the large size of the torus in these simulations, there is an ample supply of gas and magnetic fields, so the accrretion flow does not run out of gas even over the extended runtime ($2.8\times 10^5\,t_g$) of the simulations. In this sense, these simulations are more analogous to the Bondi-like simulations mentioned in Section~\ref{sec:intro}.

When postprocessing the raw simulation output, we denote time ($t$) averages of a given variable $X(t,r,\theta,\varphi)$ with an overbar: $\overline{X}$. These time averages are computed over the range $t=2\times 10^5-2.8\times 10^5\,t_g$, unless otherwise stated. For shell ($\theta,\varphi$) averages we use a bracket, $\langle X \rangle$:
\begin{equation}
    \langle X\rangle(t,r) \equiv \frac{\iint X\rho\sqrt{-g}\,d\theta\,d\varphi}{\iint \rho\sqrt{-g}\,d\theta\,d\varphi},
\end{equation}
where the integrals go over $\theta = 0-\pi$, $\varphi = 0 - 2\pi$ (the whole sphere), and $g$ is the determinant of the metric. The weighting by $\rho$ is to ensure that the shell average focuses on the disk region of the flow rather than the jet.
For the case of the density $\rho$, however, we calculate the time and shell averages slightly differently,
\begin{equation}
    \langle \overline{\rho}\rangle (r) =  \frac{\int_0^\pi\Tilde{\rho}(r,\theta)^2 \sqrt{-g}\,d\theta}{\int_0^\pi \Tilde{\rho}(r,\theta)\sqrt{-g}\,d\theta},
\end{equation}
where $\Tilde{\rho}(r,\theta)= \int_0^{2\pi} \bar{\rho}(r,\theta,\varphi) \,d\varphi/(2\pi)$ is the $t,\varphi$-averaged density. This modification is to reduce the effects of correlated fluctuations ($\rho$ is maximally correlated with itself, so it is problematic to average $\rho^2$).

We calculate the mass accretion rate as 
\begin{equation}
    \dot{M}(t,r) \equiv -\iint \rho u^r\sqrt{-g}\,d\theta d\varphi,
\end{equation}
and the energy inflow rate as
\begin{equation}
    \dot{E}(t,r) \equiv \iint T^r_t \sqrt{-g}\,d\theta\,d\varphi.
\end{equation}
We define the dimensionless feedback efficiency $\eta$ as the net energy outflow rate but excluding the rest mass energy, normalized by the rest mass energy inflow rate at $r=10\,r_g$:
\begin{equation}\label{eq:eta}
    \eta(t,r) \equiv \frac{\dot{M}(t,r)-\dot{E}(t,r)}{\dot{M}(t,10\,r_g)}.
\end{equation}
Finally, we define the dimensionless magnetic flux parameter $\phi_b$ as in \citet{Tchekhovskoy2011},
\begin{equation}\label{eq:phib}
    \phi_b(t,r)\equiv \sqrt{\frac{4\pi}{\dot{M}(t,10\,r_g)\,r_g^2}}\iint \frac{|B^r|(t,r)}{2}\sqrt{-g}\,d\theta\,d\varphi,
\end{equation}
where $B^r$ is the radial component of the magnetic field. The time-averaged versions of efficiency, $\bar{\eta}(r)$, and magnetic flux parameter, $\overline{\phi_b}(r)$, are calculated by separately time-averaging $\overline{\dot{M}}(10\,r_g)$ before including in Equations \eqref{eq:eta} and \eqref{eq:phib}, respectively.

\section{Standard Weakly Magnetized Models}\label{sec:standard models}

Torus simulations in the literature typically start with the torus quite close to the BH horizon. The radius of the pressure maximum, which we treat as the characteristic scale of the initial state, is often located at $r_{\rm max} \approx 20\,r_g$. Even the larger initial tori used in \citet{Narayan2022} had $r_{\rm max}$ only $\approx 40\,r_g$. In contrast, most of the recent Bondi-like simulations which we mentioned in Section~\ref{sec:intro} have characteristic radii, which we identify with the Bondi radius, of several $100\,r_g$ or larger.
A recent parameter study by \citet{Galishnikova2025} has shown that the variability is artificially suppressed even in Bondi-like systems when the Bondi radius is small $\sim 50\,r_g$. Therefore, we expand the torus comparable in scale to most Bondi-like simulations by setting $r_{\rm max} = 500\,r_g$ and see whether it can reproduce their strong variability.
Other than this change in the size of the initial FM torus, everything else is similar to standard torus simulations. In particular, we initialize the torus with a weak poloidal magnetic field with $\beta_{\rm max}=100$, just like in traditional torus simulations. We refer to these runs as ``standard'' torus models to distinguish them from the ``strongly magnetized'' torus models discussed in Section~\ref{sec:beta1_models}. We have run four standard torus models, summarized in Table~\ref{tab:standard models table}, corresponding to four BH spin values: $a_* = 0, ~0.5, ~0.9, ~-0.9$.

\begin{table}[h!]
  \begin{center}
    \caption{Models initialized with a large standard torus ($r_{\rm max}=500r_g, ~\beta_{\rm max}=100$) for four BH spin values $a_*$. The time-averaged (between $t=2\times10^5-2.8\times10^5t_g$) accretion rate $\overline{\dot{M}}$, dimensionless magnetic flux $\overline{\phi_b}$, and feedback efficiency $\overline{\eta}$ are listed.}
    \label{tab:standard models table}
    \begin{tabular}{l|c|c|c|c} 
    \hline
      Model Name & $a_*$ & $\overline{\dot{M}}(10\,r_g)$ & $\overline{\phi_{b}} (r_H)$ & $\overline{\eta}(10\,r_g)$ \\
      \hline
      \hline
      $a0\beta100$ & 0 & 500 & 57 & 0.030 \\
      $a.5\beta100$ & 0.5 & 320 & 64 & 0.32 \\
      $a.9\beta100$ & 0.9 & 200 & 53 & 1.2 \\
      $a-.9\beta100$ & $-0.9$ & 400 & 26 & 0.17 \\
      \hline
      \end{tabular}
  \end{center}
\end{table}

\begin{figure*}[ht!]
\gridline{
\fig{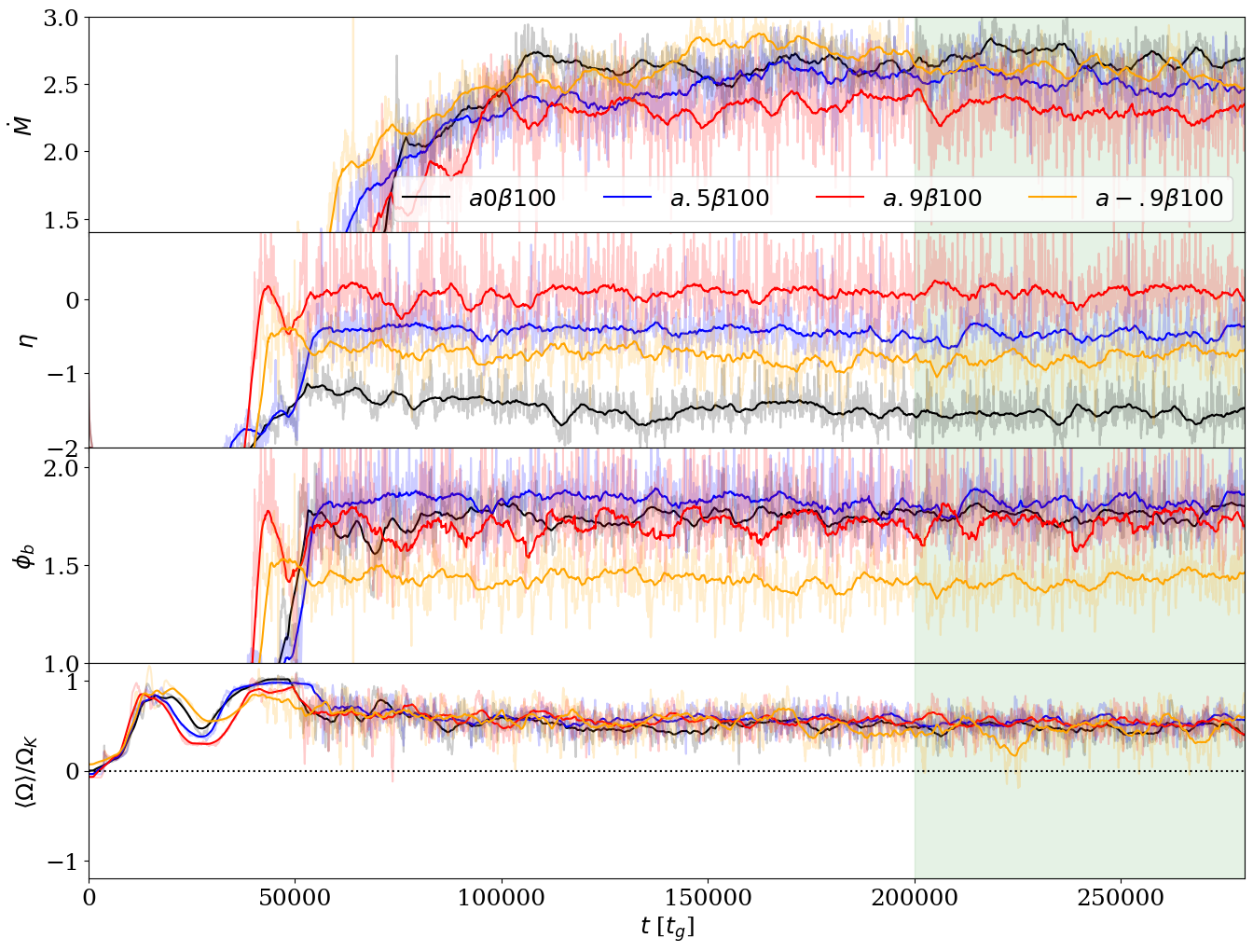}{0.99\textwidth}{}
}
\caption{
Time evolution of the mass accretion rate $\dot{M}$ at the horizon $r_H$, feedback efficiency $\eta$ at $r=10\,r_g$, dimensionless magnetic flux $\phi_b$ at $r_H$, and the shell-averaged angular velocity $\langle\Omega\rangle$ divided by the Keplerian angular velocity $\Omega_K$ at $r=10\,r_g$. The four standard torus simulations listed in \autoref{tab:standard models table}, which correspond to BH spins $a_*=0, ~0.5, ~0.9, ~-0.9$ are shown in black, blue, red, and yellow lines, respectively. For each quantity, light-colored lines show the raw output from the simulations with a cadence of $50\,t_g$, and bold lines correspond to the same data smoothed with a $5000\,t_g$ window. The time averages in \autoref{tab:standard models table} are calculated over the range $t=2\times 10^5-2.8\times 10^5\,t_g$, indicated with a light green background. Despite the substantially larger size of the tori in these simulations, the level of time-variability in the plotted quantities is similar to that seen with smaller-scale tori.
}
\label{fig:standard time evolution}
\end{figure*}

\subsection{Time Evolution}

In \autoref{fig:standard time evolution}, we show the time evolution of the accretion rate $\dot{M}(r_H)$ measured at the horizon, the feedback efficiency $\eta(10\,r_g)$ measured at $r=10\,r_g$ (to avoid numerical artifacts near the horizon), the dimensionless magnetic flux parameter $\phi_b(r_H)$, and the shell-averaged angular velocity $\langle\Omega\rangle(10\,r_g)/  \Omega_K$ scaled by the (Newtonian) Keplerian angular velocity $\Omega_K\equiv (r/r_g)^{-3/2}$. The angular velocity is $\Omega \equiv u^\varphi/u^t$. 

In contrast to traditional GRMHD simulations initialized with small initial tori, where mass begins to accrete within a few $10^3\,t_g$ and the systems reach a quasi-steady state well before $10^4\,t_g$, the present runs begin to accrete only around $t\sim 5\times10^4t_g$ and reach steady state only after $10^5\,t_g$. However, once steady state is reached, the evolution is quite similar to that seen with traditional small-scale tori. Specifically, $\eta$, $\phi_b$ and $\langle\Omega\rangle$ all exhibit steady behavior, with relatively modest fluctuations around a well-defined time-averaged mean.

One notable difference from previous work is that the $\dot{M}$ profiles in \autoref{fig:standard time evolution} show no secular decline with inreasing time. This is because the large initial tori used in this work contain a huge reservoir of gas which is hardly depleted even by $t=2.8\times10^5t_g$. Note the contrast with Figure~2 in \citet{Narayan2022}, and especially Figure~4 in \citet{Chatterjee2022}, where those smaller tori experience significant mass depletion and a noticeable decline of $\dot{M}$ with time.

With regard to our motivation to understand the intermittent behavior of jet activity, it is noteworth that there is no evidence of wild swings in $\eta$ or $\phi_b$ in the present simulations. Even though the characteristic scale of these torus simulations ($r_{\rm max}=500r_g$) is comparable to
the Bondi radii of several recent Bondi-like simulations, these standard tori do not reproduce the intermittent jets seen in the Bondi-like simulations. This indicates that intermittent jets are not caused purely as a result of having a large dynamic range of scales. Other factors must be important as well.

\subsection{Consistency with \citet{Blandford1977}}

\begin{figure}[ht!]
\gridline{
\fig{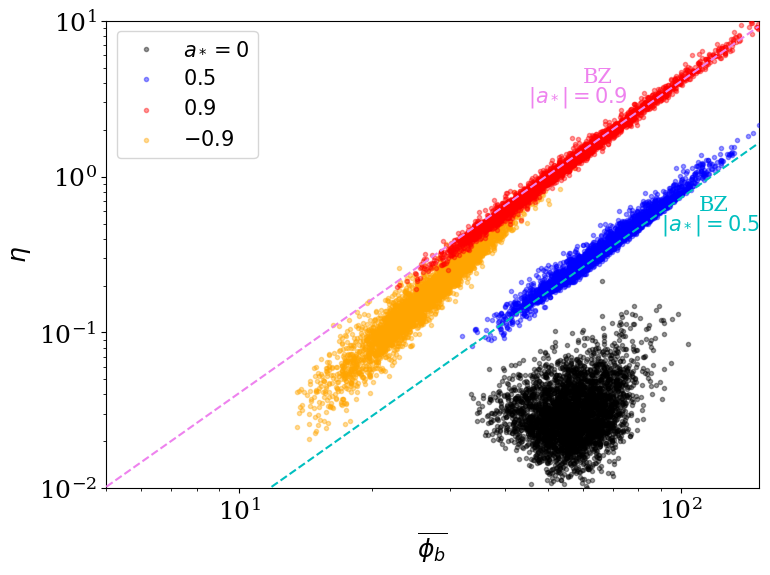}{0.5\textwidth}{}
}
\caption{Correlation between the efficiency $\eta(t,10\,r_g)$ and the magnetic flux parameter $\phi_{b}(t,r_H)$ for the four standard weakly magnetized simulations listed in \autoref{tab:standard models table}. Black, blue, red, yellow dots correspond to spins $a_*=0$, $0.5$, $0.9$, $-0.9$, respectively. The simulation data show excellent agreement with the $\eta-\phi_b$ relation predicted the BZ theory (\autoref{eq:BZ}), shown as cyan and pink dashed lines for spins $|a_*|=0.5$ and $0.9$, respectively.
}\label{fig:standard eta vs phi}
\end{figure}

The time-dependent efficiency $\eta$ and magnetic flux parameter $\phi_b$ during the period $t\geq 10^5\,t_g$ (when accretion has reached steady state) are plotted for the four standard models in \autoref{fig:standard eta vs phi}. 
The BZ jet efficiency formula from \citet{Tchekhovskoy2010,Tchekhovskoy2011} is
\begin{equation}\label{eq:BZ}
    \eta(a_*,\phi_b) \approx \frac{k}{4\pi} \phi_b^2\left(\Omega_H r_g\right)^2 (1+1.38\Omega_H^2r_g^2 - 9.2\Omega_H^4r_g^4),
\end{equation}
where $\Omega_H = a_*/(2r_H)$ is the horizon angular velocity and the constant $k$ depends weakly on the geometry of the magnetic field: $k=1/(6\pi)\approx 0.053$ for a monopolar field, and $k\approx 0.044$ for a parabolic field. The BZ relation with $k=0.05$ is plotted in \autoref{fig:standard eta vs phi} for $|a_*|=0.5$ and $|a_*|=0.9$ in cyan and pink dashed lines, respectively.

We find that the efficiency $\eta$ is tightly correlated with $\phi_b$ and closely follows the BZ formula (\autoref{eq:BZ}) for the $a_*=0.5$ and 0.9 models. There can be little doubt that the energy outflow in these two models is primarily via the BZ mechanism, i.e., the power originates in the BH spin. The retrograde model $a_*=-0.9$ deviates from the theoretical BZ relation at lower values of the  magnetic flux parameter $\phi_b\lesssim 20$; perhaps it suggests that  a minimum magnetic flux $\phi_b$ is required for the BZ process to be effective. In the case of the $a_*=0$ model, there is no BH spin energy and the BZ mechanism does not operate. The energy outflow here is likely powered by reconnection near the BH \citep{Cho2023,Cho2024}.

\subsection{Time-averaged Fluxes}\label{sec:density profile}

The time-averaged accretion rate $\overline{\dot{M}}$, magnetic flux parameter $\overline{\phi_b}$, and jet efficiency $\overline{\eta}$ for our four standard torus models are listed in \autoref{tab:standard models table}. The estimated values of $\overline{\phi_b}$ and $\overline{\eta}$ are similar to those reported for the same spin values in \citet[][see their Table~3]{Narayan2022}. It is very reassuring that $\overline{\eta}$, in particular, which is an important parameter that measures jet power and BH feedback, is insensitive to
the size of the initial torus. 

With regard to the time-averaged mass accretion rate $\overline{\dot{M}}$, \autoref{tab:standard models table} indicates that this quantity has an interesting dependence on the BH spin. The accretion rates in the  $a_*=0.5$ and $a_*=0.9$ models are $\sim 1.6$ and $\sim 2.5$ times smaller, respectively, than the accretion rate for a non-spinning BH, $a_*=0$. This is a meaningful cross-comparison because all the simulations use an identical initial torus. Furthermore, because of the large mass in the torus there is negligible mass loss or secular evolution of the torus. Thus, the different simulations have virtually identical boundary conditions near $r \sim r_{\rm max}$. Any differences in $\overline{\dot{M}}$ must be induced by some effect associated with the BH spin, though exactly how it works is presently unclear.

\begin{figure}[]
\gridline{
\fig{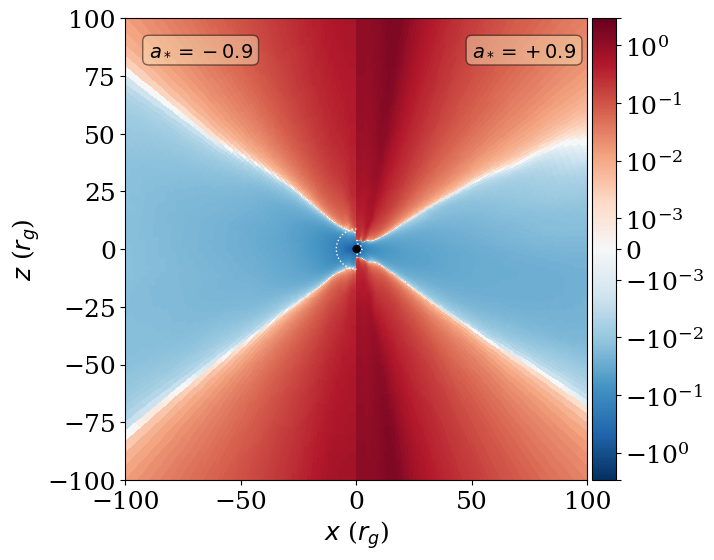}{0.5\textwidth}{}
            } 
\caption{Poloidal slice of $(t,\varphi)$-averaged $u^r$ of the (\textit{left half}) retrograde $a_*=-0.9$ (run $a-0.9\beta100$)  and (\textit{right half}) prograde $a_*=0.9$ (run $a.9\beta100$) models. Inflowing (outflowing) regions are shown in blue (red). Although both models have the same magnitude of the BH spin, the inflowing $\theta$ range is wider for the retrograde $a_*=-0.9$ run, possibly because of the weaker feedback efficiency $\eta$ in this model. This geometrical effect might explain why this model has a higher accretion rate $\dot{M}$ than the $a_*=0.9$ model. The ISCO radii $r_{\rm ISCO}$ of the two models are shown as white dotted half-circles.
}\label{fig:tavged_ur}
\end{figure}

It could be that the stronger jet feedback associated with rapid BH spin narrows the range of $\theta$ over which gas flows in and mildly suppresses the mass accretion rate by a factor of a few. The density-weighted $t,\varphi$-averaged $u^r$ distribution is compared between prograde $a_*=0.9$ and retrograde $a_*=-0.9$ runs in Figure~\ref{fig:tavged_ur}. The innermost stable circular orbit (ISCO) is also shown for reference. The boundary between inflow and outflow is located at different angles in the two models. The retrograde $a_*=-0.9$ run, with its weaker feedback, allows a wider solid angle for gas inflow compared to the prograde $a_*=0.9$ model. Thus, a factor of a few difference in accretion rates might be explained by such geometrical effects shaped by the jet strength.

Recent GRMHD simulations of Bondi-like accretion find that the mass accretion rate on the BH, $\dot{M}$, is suppressed with respect to the classic Bondi mass accretion rate, $\dot{M}_B$, by a factor which depends on the ratio of the Bondi radius $r_B$ to $r_g$ \citep{Cho2024,Guo2025}. We find here that the suppression factor has an additional weak dependence on BH spin (\autoref{tab:standard models table} and \autoref{tab:beta1 models table}). A similar dependence of $\dot{M}$ on BH spin $a_*$ was shown for a small FM torus by \citet{Guo2025}, but its limited gas supply resulted in a steady decline in the accretion rate. Here, we confirm the result with a better measurement of the accretion rate $\dot{M}$ as the abundant gas in the large torus helps maintaining a roughly constant accretion rate $\dot{M}$ over time.

\subsection{Relativistic Jet Velocity}

\begin{figure}[ht!]
\gridline{
\fig{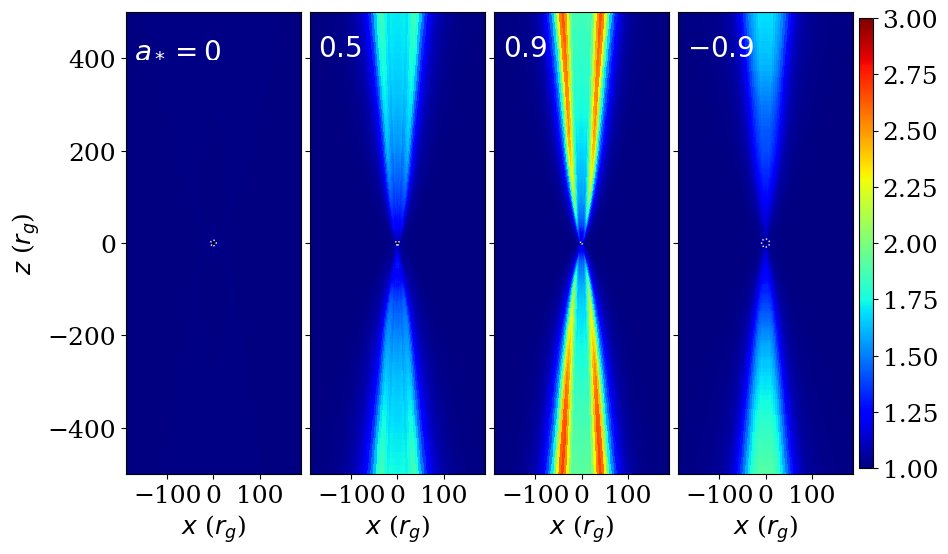}{0.48\textwidth}{}
}   
\caption{Distribution of the time- and $\varphi$-averaged Lorentz factor $\Gamma$ in the poloidal plane for the four standard simulations (from left to right): $a0\beta100$ ($a_*=0$), $a.5\beta100$ ($a_*=0.5$), $a.9\beta100$ ($a_*=0.9$), and $a-.9\beta100$ ($a_*=-0.9$).}\label{fig:standard gamma}
\end{figure}

\autoref{fig:standard gamma} shows the poloidal distribution of the density-weighted, $t,\varphi$-averaged Lorentz factor $\Gamma$, measured in the Eulerian frame, for the four standard torus simulations. It is clear that the jet $\Gamma$, just like the jet efficiency $\eta$, has a strong dependence on BH spin. Note that $\phi_{b}$ is roughly the same for the $a_* = 0, ~0.5, ~0.9$ models, so the large differences in their $\Gamma$ distribution are primarily coming from the BH angular velocity factor $\Omega_H^2$ in \autoref{eq:BZ}. The jet in the retrograde model, $a_*=-0.9$, is less relativistic than in the corresponding prograde model, $a_*=0.9$. In this case, $\Omega_H^2$ is the same for the two models, so the diffference in $\Gamma$ most likely arises from the smaller magnetic flux $\phi_b$ in the retrograde model.

\subsection{Slope of the Density Profile}

\begin{figure*}[ht!]
\gridline{
\fig{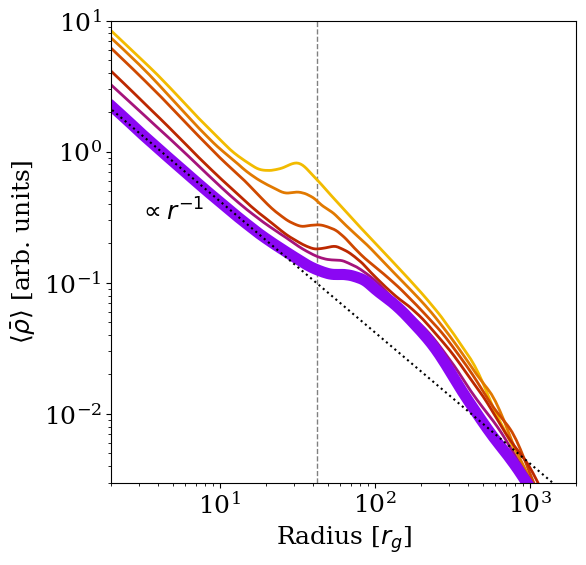}{0.33\textwidth}{(a) Small torus $a_*=0$}
\fig{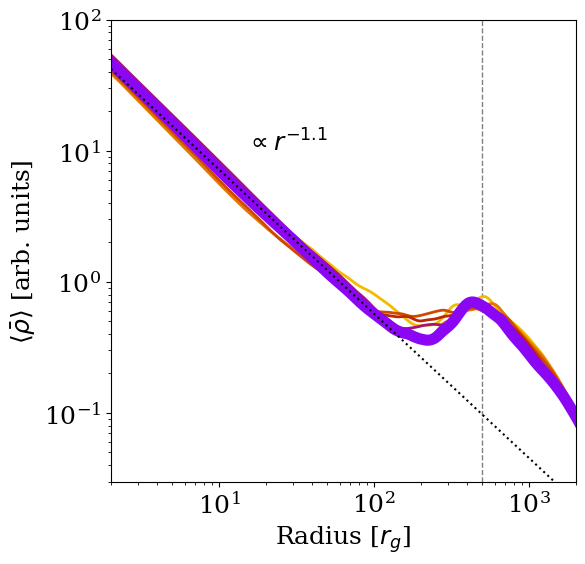}{0.33\textwidth}{(b) Large torus $a_*=0$}
\fig{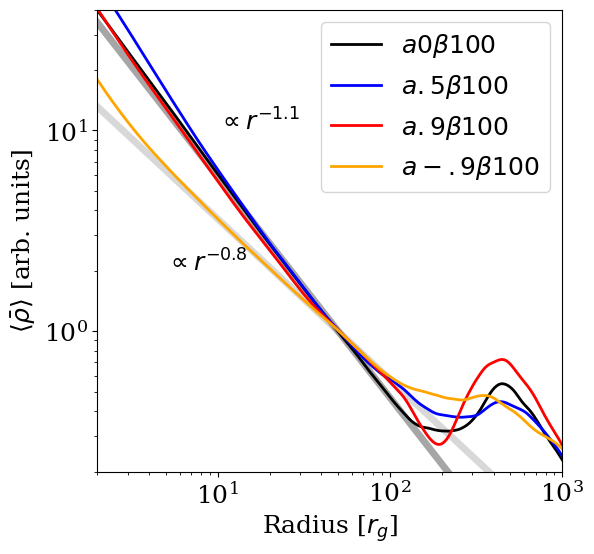}{0.33\textwidth}{(c) Large tori final density profile}
            }   
\caption{Density profiles in torus simulations. (\emph{a}) Radial profiles of the $t,\theta,\varphi$-averaged density $\langle\bar{\rho}\rangle(r)$ in the $a_*=0$ small torus (pressure maximum $r_{\rm max}$ at $42.43\,r_g$ shown as a dashed vertical line) simulation described in \citet{Narayan2022}, time-averaged over 6 time chunks split between $t=9\times 10^3-10^5\,t_g$. Time progresses from yellow to purple, with the profile from the final time-average highlighted in bold purple. The dotted line corresponds to a slope of $-1$. (\emph{b}) Same as panel (a) but for the large torus model ($r_{\rm max} = 500\,r_g$, vertical dashed line) with zero BH spin (model $a0\beta100$) described in the present paper. The time-averages are over 6 time chunks split between $t=10^5-2.8\times 10^5\,t_g$ from yellow to purple. Because the initial torus is located at a much larger radius, the density profile can be reliably followed over a larger range of radius in this model, and found to scale as  $\rho\propto r^{-1.1}$, shown an a black dotted line. (\emph{c}) The density profiles of all four standard torus models (\autoref{tab:standard models table}), each normalized by its density at $50\,r_g$. The dark and light gray lines show density scalings of $\rho(r)\propto r^{-1.1}$ and  $\rho(r)\propto r^{-0.8}$, respectively.
}\label{fig:density_comparison}
\end{figure*}

Several recent Bondi-like simulation studies have reported that the radial density profile of the steady state accretion flow scales roughly as $\rho \propto r^{-1}$ \citep{Ressler2021,Guo2023,Cho2023,Cho2024,Guo2025}. A slope of $-1$ is in between $-3/2$ predicted by the original ADAF model \citep{Narayan1994,Narayan1995}, and $-1/2$ predicted by models based on convection \citep{Narayan2000,Quataert2000}. \citet{Xu2023} has proposed an explanation for the in-between slope based on hydrodynamic turbulence from convection.

There is very little previous work on the density profile in GRMHD simulations starting from a FM torus (with a few exceptions e.g., \citealt{Begelman2022,Chatterjee2022}). A major problem is that, in most work, the steady state region of the accretion flow extends over only a small range of radius. \autoref{fig:density_comparison}(a) shows the density profile averaged over 6 different time chunks in the $a_*=0$ simulation reported in \citet{Narayan2022}. Even though this was a moderately large-scale simulation, with torus initial inner edge at $r_{\rm in}=20\,r_g$ and pressure maximum at $r_{\rm max}=42.43\,r_g$, it is clear that the density profile is strongly affected by the presence of the torus. At early times, there is a bump in the density profile at the radius of the pressure maximum $r_{\rm max}$, and even by the end of the simulation, a bump survives at $r\sim 70\,r_g$. When attempting to fit a power-law to the final profile (thick purple line), one obtains $\rho\propto r^{-1}$ between $r=r_H=2\,r_g$ and $\sim 20\,r_g$, followed by a shallower slope in the transition region between $20r_g$ and the inner edge of the bump. It is hard to measure a reliable slope when the data are limited to such a small radius range, that too near the inner boundary.

The new standard torus simulations described in this section are more suited for measuring the slope of the density profile since the torus here is initially located much further away, leaving a substantial range of radius to study the density scaling.
The time-averaged density profiles, corresponding to 6 time chunks split between $10^5-2.8\times 10^5\,t_g$, for the non-spinning $a_*=0$ model ($a0\beta100$) are shown in \autoref{fig:density_comparison}(b). Compared to the small torus model in panel (a), we see that there is an excellent power-law dependence of the density out to $r\sim100\,r_g$. The slope is measured to be $-1.1$ (similar to the value found by \citealt{Chatterjee2022}), ruling out a slope of $-1$. \citet{Xu2023}'s analytical model predicts a slope of $-0.8$ or even shallower, which is inconsistent with the present simulation data. This might be because their analytical model is based on hydrodynamic convection, whereas the present MAD simulation is dominated by magnetic field dynamics.

\autoref{fig:density_comparison}(c) shows the final density profiles of all four models. The models with $a_* = 0, ~0.5, ~0.9$ all exhibit density scaling of $\rho(r)\propto r^{-1.1}$, or perhaps slightly steeper. However, the retrograde model, $a_*=-0.9$, exhibits a shallower slope of $\rho(r)\propto r^{-0.8}$. Curiously, this is the slope found by \citet{Chatterjee2022} for the SANE state, although this may be coincidental.

\section{Strongly Magnetized Models}\label{sec:beta1_models}

\begin{figure*}[ht!]
\gridline{
\fig{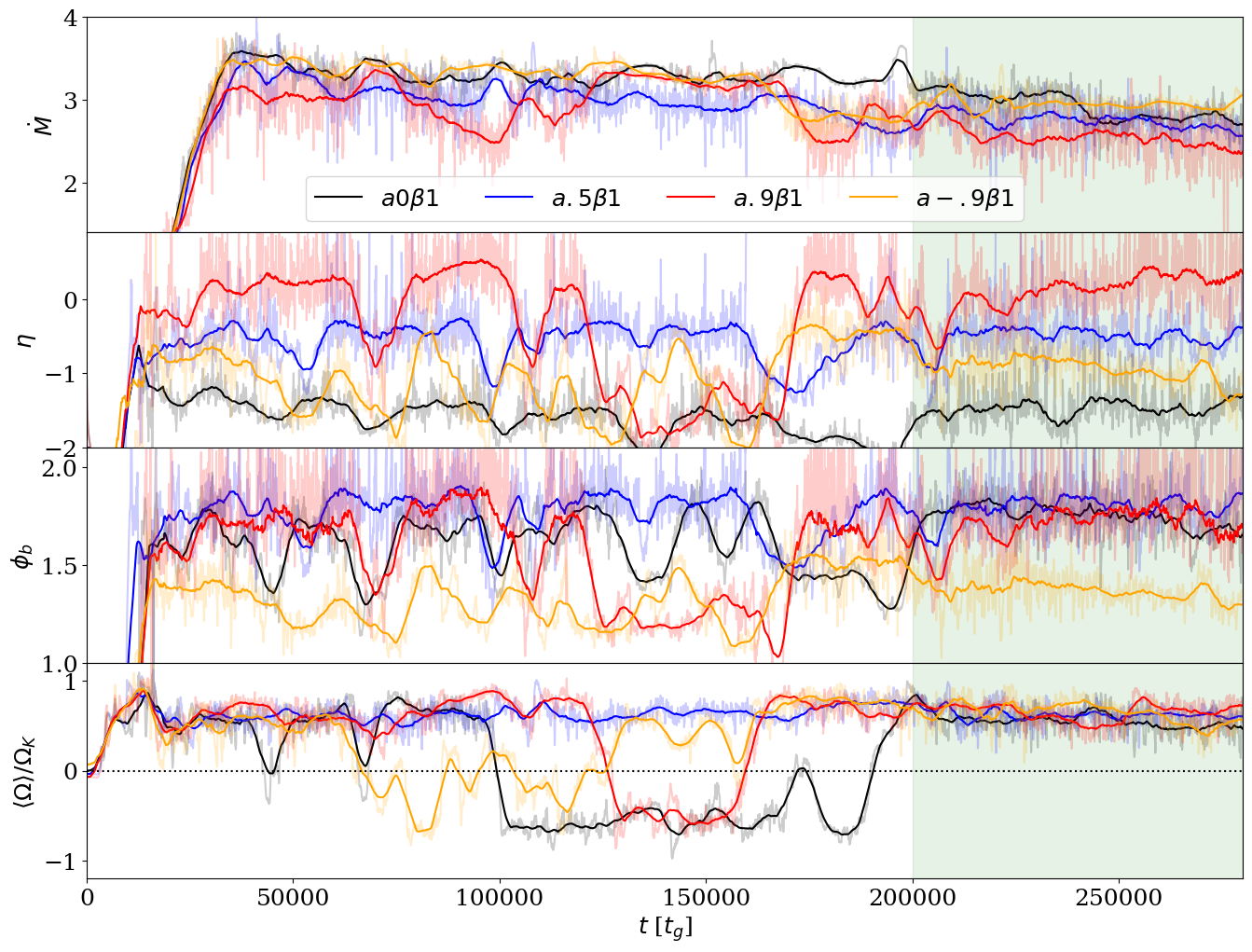}{0.99\textwidth}{}}
\caption{
Same as \autoref{fig:standard time evolution}, but for the strongly magnetized torus models with $\beta_{\rm max}\sim 1$. The variability in feedback efficiency $\eta$ and dimensionless magnetic flux $\phi_b$ is significantly larger, and the angular velocity $\Omega$ even flips its sign during certain periods.
}\label{fig:strongly magnetized time evolution}
\end{figure*}

The large ``standard torus'' models in Section~\ref{sec:standard models} showed neither of the two effects we are seeking to understand: intermittent jet activity, and flips in the angular velocity. To further explore the problem, we have run an equivalent series of four ``strongly magnetized'' models. In these simulations, we increase the initial poloidal magnetic field strength substantially, by setting $\beta_{\rm max}=1$, while keeping other setup details essentially the same. We again consider four BH spins, $a_* = 0, ~0.5, ~0.9, ~-0.9$, as summarized in \autoref{tab:beta1 models table}.

\begin{table}[h!]
  \begin{center}
    \caption{Same as \autoref{tab:standard models table}, but for models initialized with a strongly magnetized torus ($\beta_{\rm max}=1$).}
    \label{tab:beta1 models table}
    \begin{tabular}{l|c|c|c|c} 
    \hline
      Model Name & $a_*$ & $\overline{\dot{M}}(10\,r_g)$ & $\overline{\phi_{b}} (r_H)$ & $\overline{\eta}(10\,r_g)$ \\
      \hline
      \hline
      $a0\beta1$ & 0 & 890 & 56 & 0.035 \\
      $a.5\beta1$ & 0.5 & 570 & 60 & 0.30 \\
      $a.9\beta1$ & 0.9 & 370 & 53 & 1.2 \\
      $a-.9\beta1$ & -0.9 & 920 & 23 & 0.13 \\
      \hline
      \end{tabular}
  \end{center}
\end{table}

\subsection{Time Evolution Now Shows Large Fluctuations}\label{sec:beta1_evolution}
The time evolution of $\dot{M}(r_H)$, $\eta(10r_g)$, $\phi_b(r_H)$, and $\langle\Omega\rangle/\Omega_K (10\,r_g)$ for the four strongly magnetized tori are shown in \autoref{fig:strongly magnetized time evolution}. A comparison with the equivalent results for standard tori in \autoref{fig:standard time evolution} shows several notable differences.

The present models evolve significantly more quickly than the models in Section~\ref{sec:standard models}. This is not surprising since the stronger initial field strength drives the systems to the MAD state almost instantly, thereby initiating rapid angular momentum removal, whereas the standard tori have to wait until the field strength grows (via the MRI) to the MAD level before reaching an equivalent state.

The time evolution of $\eta$ and $\phi_b$ is significantly more chaotic in the present strongly magnetized models (\autoref{fig:strongly magnetized time evolution}) compared to the standard models (\autoref{fig:standard time evolution}). Notice in particular the very large fluctuations in the feedback efficiency $\eta$ in the $a_*=0.9$ model (red curve), which shows factor $\sim1000$ swings in the raw data and factor $\sim100$ fluctuations even after smoothing the data with a $5000\,t_g$ window function. The level of fluctuations is similar to what has been found in previous Bondi-like simulations \citep{Ressler2021,Kwan2023,Lalakos2024, Galishnikova2025,Kim2025,Guo2025,Lalakos2025}, but we now see it also in our large torus simulations (it has been seen also by \citealt{Chan2025}, who ran modified small-scale FM tori). The magnetic flux $\phi_b$ similarly shows large variations in our simulations which are strongly correlated with the variations in $\eta$ (explored further in \autoref{fig:strongly magnetized eta vs phi}).

Next, the angular velocity $\langle\Omega\rangle (10\,r_g)$ shows large fluctuations in \autoref{fig:strongly magnetized time evolution}, including episodes of a complete reversal in the sense of rotation (sign flip in $\langle\Omega\rangle/\Omega_K$). This is particularly evident in the $a_*=0$ (black curve) and $a_*=0.9$ (red) models. A similar flipping of the angular velocity was observed by \citet{Cho2024} in highly magnetized Bondi-like accretion on non-spinning BHs and by \citet{Galishnikova2025} in weakly rotating Bondi-like accretion on spinning BHs.
Nothing equivalent has been reported in any of the other Bondi-like simulations with spinning BHs. Here we show that even rapidly rotating FM torus models show 
$\langle\Omega\rangle$ flips for both spinning and non-spinning BHs. The connection between rotation flips and the wildly fluctuating efficiency $\eta$ will be discussed in depth in Section~\ref{sec:eta_variation}. 

Interestingly, we see both large fluctuations in $\eta$, $\phi_b$ and sign flips in $\langle\Omega\rangle$ only when the torus is initialized with a strong magnetic field ($\beta_{\rm max}=1$, this Section) and not with weak magnetic field ($\beta_{\rm max}=100$, Section~\ref{sec:standard models}). As suggested by \citet{Cho2024}, the strong field probably brakes the orbiting gas strongly and scrambles the rotation. The same ``magnetic braking'' potentially can explain also the switching rotation in \citet{Pen2003}'s Newtonian MHD simulation and steadily decreasing $\Omega/\Omega_K$ over time in \citet{Narayan2012}'s MAD simulation.

\begin{figure*}[ht!]
\gridline{
\fig{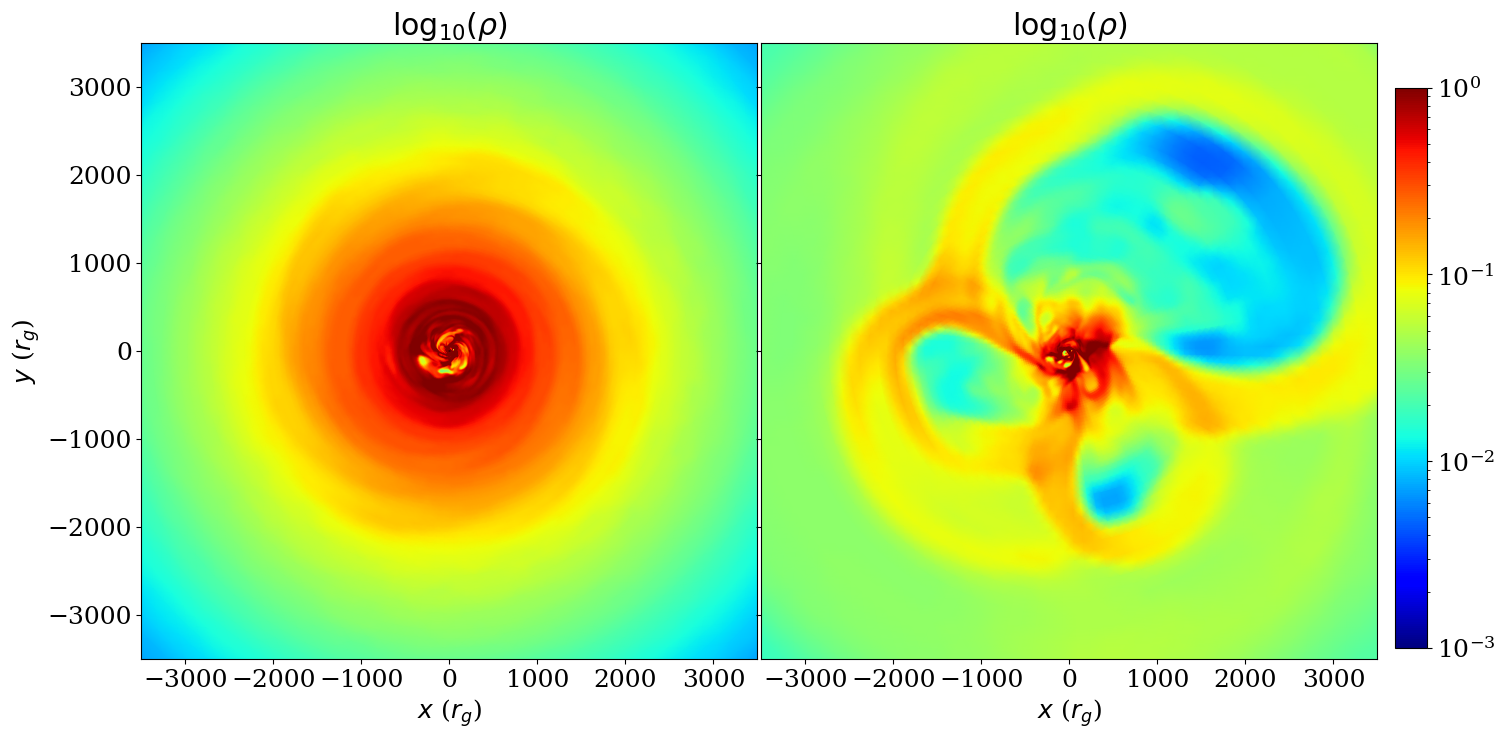}{0.99\textwidth}{}
 }
\caption{Equatorial slices at the last timestep, $t=2.8\times10^5\,t_g$, of $a_*=0$ BH simulations with (\textit{left}) weakly magnetized ($\beta\sim100$, model $a0\beta100$) and (\textit{right}) strongly magnetized ($\beta\sim1$, model $a0\beta1$) tori. The flux eruptions,  distinguished as low density regions, reach farther out, up to $\sim 3000\,r_g$, in the strongly magnetized model compared to $r\lesssim 200\,r_g$ in the standard weakly magnetized model.}\label{fig:comparison_snapshot} 
\end{figure*}

In addition to the strong variability in $\eta$, $\phi_b$, $\langle\Omega\rangle$ in the strongly magnetized models, there is also a striking difference in the structure of their accretion flows compared to the standard weakly magnetized models. \autoref{fig:comparison_snapshot} shows the final snapshots ($t=2.8\times 10^5\,t_g$) of the equatorial plane of the non-spinning $a_*=0$ BH simulations of the standard $\beta_{\rm max}=100$ (left) and highly magnetized $\beta_{\rm max}=1$ (right) tori. The standard torus has magnetic flux bundles, traced by regions of low density, confined to small radii $\lesssim 200\,r_g$, whereas in the strong initial magnetic field torus, the flux eruptions have reached an order of magnitude farther out in radius, to $r\sim 3000\,r_g$. This difference could be due to several reasons. As already mentioned, the strong initial magnetic field accelerates the onset of accretion onto the BH, so early-formed outflowing flux bundles have more time to propagate out and move to larger radii. Another explanation is that, since coherent rotation survives in the weakly magnetized torus, it suppresses the buoyant/convective outward motion of flux bundles. On the other hand, for the strongly magnetized torus, the stronger field powers more energetic flux eruptions and there is less rotational shear to stop them from moving out.

The larger bubbles of low density gas in the strongly magnetized torus, and the correspondingly more rapid outflow of flux bundles, might be responsible for the secular decrease in the accretion rate $\dot{M}$ seen in \autoref{fig:strongly magnetized time evolution}. The large-scale flux eruptions re-distribute gas density near the BH, which results in stronger variations in the accretion rate. As the characteristic size of the flux bundles is $\sim 3000\,r_g$, the associated timescale is of order $\sim 10^5\,t_g$, comparable to the total duration of the simulation. If the simulations are continued longer, e.g., using a numerical acceleration scheme like the multi-zone \citep{Cho2023,Cho2024} or cyclic-zoom \citep{Guo2025} methods, the system might eventually relax to a quasisteady state.

\subsection{Time-averaged Profiles}
Although our primary interest is the large time variability we find in the strongly magnetized tori, for completeness we also list the time-averaged values of $\overline{\dot{M}}(10\,r_g)$, $\overline{\phi_b}(r_H)$, and $\overline{\eta}(10\,r_g)$ in \autoref{tab:beta1 models table}.

\begin{figure}[ht!]
\gridline{
\fig{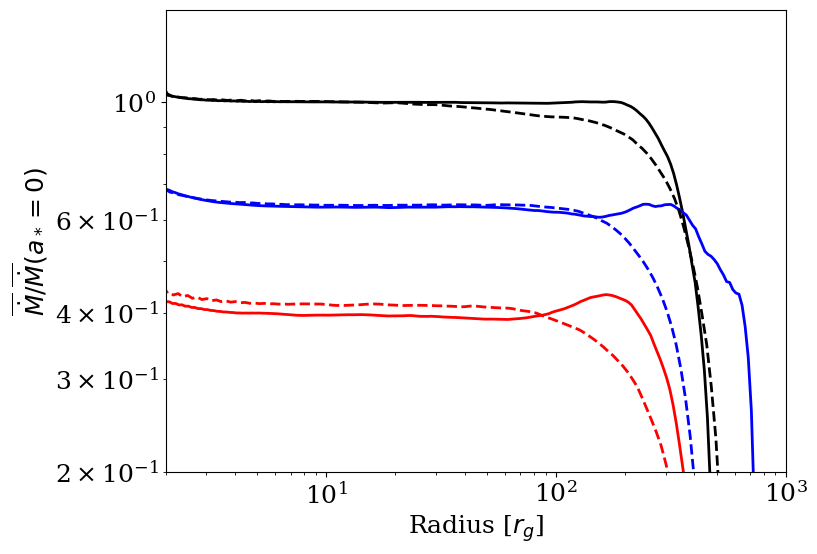}{0.45\textwidth}{}
            }   
\caption{Dependence of the accretion rate $\dot{M}$ on the BH spin $a_*$. The accretion rate profiles $\overline{\dot{M}}(r)$ for $a_*=0$ (black), $a_*=0.5$ (blue), and $a_*=0.9$ (red) are normalized by the accretion rate of the non-spinning BH ($a_*=0$). The solid lines are for the standard ($\beta_{\rm max}=100$) runs and the dashed lines are for the strongly magnetized ($\beta_{\rm max}=1$) runs.
}\label{fig:Mdot_comparison}
\end{figure}

The accretion rates $\overline{\dot{M}}(10r_g)$, time-averaged over a relatively quiescent phase, $t=2.-2.8\times 10^5\,t_g$, show a modest dependence on BH spin, similar to that seen in the standard tori (\autoref{tab:standard models table}). The dashed lines in \autoref{fig:Mdot_comparison} are the accretion rate profiles $\dot{M}(r)$ of the strongly magnetized tori normalized by $\overline{\dot{M}}(10r_g)$ of the $a_*=0$ model; the corresponding results for the standard torus runs are shown as solid lines. In both sets of models, the accretion rate for spinning BHs is suppressed compared to the non-spinning BH model. The suppression factor is $\sim 1.6$ and $\sim 2.4$, respectively, for BH spin $a_*=0.5$ and $a_*=0.9$. The time-averaged dimensionless magnetic flux $\phi_b(r_H)$ and feedback efficiency $\eta(10\,r_g)$ values in Table~\ref{tab:beta1 models table} are also similar to those found for standard tori with the same BH spin (Table~\ref{tab:standard models table}).

\begin{figure}[]
\gridline{
\fig{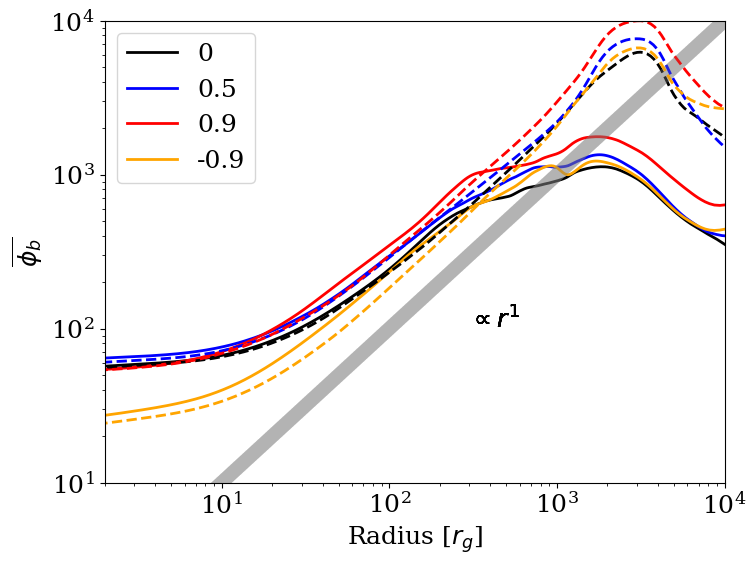}{0.45\textwidth}{}
            }   
\caption{Time averaged radial profiles of the dimensionless magnetic flux parameter $\phi_b(r)$ for all the torus simulations. Solid (dashed) lines show the profiles for runs with initially weak (strong) magnetic fields. Different colors represent different BH spins as labeled. All four strongly magnetized models exhibit self-similar $\phi_b\propto r^1$ profiles over a wide radius range up to $r\sim 3000\,r_g$, indicative of a radially extended MAD state. The initially weakly magnetized models have their MAD zones truncated around $r\approx200r_g$.
}\label{fig:phib_comparison}
\end{figure}

While the magnetic flux parameter $\phi_b (r_H)$ at the horizon is similar in standard and strongly magnetized models, a comparison of the radial profiles of $\phi_b(r)$ shows differences (\autoref{fig:phib_comparison}). Both sets of models have $\phi_b(r) \sim$ constant for $r \lesssim 10\,r_g$, indicative of a constant magnetic flux, and a switch to $\phi_b(r)\propto r^1$ (flux increasing with radius) beyond $r\sim 10\,r_g$. Following the logic presented in \citet{Cho2023} and in Section~\ref{sec:phib_analytic}, the latter scaling  implies a magnetically saturated state with $\beta\sim 1$, corresponding to a radially extended MAD flow. The extent of this MAD zone is, however, quite different in the two sets of models. In the standard tori, the slope of $\phi_b(r)$ begins to flatten at radii $r\gtrsim 200\,r_g$, whereas in the strongly magnetized tori the $\propto r^1$ scaling continues up to $\sim 3000\,r_g$, implying that a saturated MAD state is achieved over a significantly more extended range of radius. The radial extent of magnetic saturation is consistent with the bubble sizes for both standard and strongly magnetized tori in \autoref{fig:comparison_snapshot}. These differences appear to be related to the choice of initial $\beta_{\rm max}$ in the two sets of models. With their weaker initial magnetic field, standard tori ($\beta_{\rm max}=100$) have less total magnetic flux available, and are able to push the system into the MAD state over only a modest range of radius. The strongly magnetized tori ($\beta_{\rm max}=1$) have a factor of 10 larger magnetic flux and thus are able to produce a MAD flow extending out to a factor $\sim10$ larger radius over the same simulation time.

\subsection{Why is the magnetic flux $\phi_b$ the way it is?}\label{sec:phib_analytic}
All magnetic flux profiles $\phi_b(r)$ in \autoref{fig:phib_comparison} exhibit a common characteristic shape, where the slope changes from $\phi_b(r)\sim ({\rm constant})$ to $\phi_b(r)\propto r^1$ at around radius $r\sim 10\,r_g$. This has also been seen in the multi-zone simulations of \citet{Cho2023}, where the same radius $\sim 10\,r_g$ is where the magnetic fields change their geometry from monopolar to vertical (e.g., top left panel of Figure 2 in \citealt{Cho2023}) and the plasma-$\beta$ changes character from $\beta\ll 1$ to $\beta\sim 1$ (e.g., Figure 1(e) in \citealt{Cho2023}).

The change in the character of $\phi_b(r)$ at $r\sim 10\,r_g$ is indicative of a change in the magnetic field geometry.
A split-monopolar field  has $B^r(r)\propto r^{-2}$ and has the same number of field lines threading spheres of different radii. This gives $\phi_b(r) \sim$ constant, and describes the behavior at $r\lesssim 10\,r_g$. However, for a vertical field, new field lines will be included with increasing radius, leading to an increasing flux $\phi_b(r)$ with increasing radius. From \autoref{eq:phib},  $\phi_b(r)\propto r^1$ implies that $B^r(r)\propto r^{-1}$.

We can now ask why there is an overall qualitative change at $r\sim10\,r_g$. Near the BH, magnetic pressure dominates over gas pressure $p_b\gg p_g$, or equivalently $\beta\ll 1$, likely because accreted magnetic field lines pile up near the horizon while gas continues to disappear through the horizon. With monopolar magnetic field geometry, the magnetic pressure in the vicinity of the BH is then $p_b\propto r^{-4}$. In the meantime, the gas pressure is $p_g\propto r^{-2}$ because $\rho\propto r^{-1}$ and $T\propto r^{-1}$ from \citet{Cho2023,Cho2024}. Therefore, at some distance from the BH, the gas pressure $p_g$ will begin to balance the magnetic pressure $p_b$; this is analogous to the physics change near the magnetospheric radius (e.g., \citealt{Frank2002}) of neutron stars (X-ray pulsars). We check whether the radius of equipartition, $p_g\approx p_b$, which we call the magnetospheric radius $r_M$, is indeed $\sim 10\,r_g$ where many magnetic properties (slope of $\phi_b(r)$, field geometry, slope of plasma-$\beta(r)$) switch behavior simultaneously in our MAD simulations.

To calculate the normalization for the magnetic pressure $p_b$, we use the saturated magnetic flux at the horizon, typically $\phi_{b,0} \approx 20-60$  \citep{Tchekhovskoy2012,Narayan2022}, where the subscript $0$ indicates the measured value at the horizon hereafter. Assuming infall at the speed of light at the horizon and spherical symmetry, the mass accretion rate is $\dot{M}_0\approx 4\pi r_H^2\rho_0$, where $\rho_0$ is the density at the horizon. Also assuming that $B^r_0$ is independent of $\theta,\varphi$, the magnetic flux at the horizon  is
\begin{align*}
    \phi_{b,0}&=\left.\sqrt{\frac{4\pi}{\dot{M}r_g^2}}\iint \frac{|B^r|}{2}\sqrt{-g}\,d\theta\,d\varphi\right\vert_{r=r_H}\\
    &\approx\sqrt{\frac{4\pi}{4\pi r_H^2 \rho_0 r_g^2}} 2\pi r_H^2 |B^r_{0}|.
\end{align*}
Then, solving for $B^r_0$, we obtain
\begin{equation*}
    |B^r_0|=\frac{\phi_{b,0}\sqrt{\rho_0}\,r_g}{2\pi r_H},
\end{equation*}
\begin{equation*}
    |B^r|(r) = |B^r_0|\left(\frac{r}{r_H}\right)^{-2},
\end{equation*}
where we have introduced the radial scaling $B^r \propto r^{-2}$ of the split-monopole field geometry.
The resulting magnetic pressure is
\begin{equation*}
    p_b = \frac{\phi_{b,0}^2\rho_0 r_g^2}{8\pi^2 r_H^2}\left(\frac{r}{r_H}\right)^{-4}.
\end{equation*}
Meanwhile, the gas pressure is
\begin{align*}
    p_g = \rho T =\rho_0 T_0 \left(\frac{r}{r_H}\right)^{-2},
\end{align*}
with $T_0$ representing the temperature at the horizon in the power-law scaling.
Hence, the magnetospheric radius where the two pressures balance is 
\begin{equation}\label{eq:rm_phib}
    r_M = \frac{\phi_{b,0}}{\sqrt{8\pi^2 T_0}}\, r_g,
\end{equation}
which only depends on the horizon magnetic flux $\phi_{b,0}$ and the temperature normalization $T_0$.

In \citet{Cho2023,Cho2024}, the horizon magnetic flux $\phi_b$ saturates around $\phi_{b,0} \sim 30$ and the temperature normalization is $T_0\approx 0.1$. Then, the magnetospheric radius is expected to be at
\begin{equation*}
    r_M\approx 10.7 \, r_g,
\end{equation*}
which is consistent with where the $\phi_b(r)$ slope changes in \autoref{fig:phib_comparison}.
Inserting a different $\phi_{b,0}=20-50$ only changes the magnetospheric radius $r_M$ by a factor of order unity.

Although \autoref{eq:rm_phib} quantitatively relates the magnetospheric radius $r_M$ with the magnetic flux at the horizon $\phi_{b,0}$, there is still the question of how the level of saturated magnetic flux $\phi_{b,0}$ itself is determined. An interesting dependence of horizon flux $\phi_{b,0}$ on the BH spin $a_*$ was revealed in \citet{Tchekhovskoy2012,Narayan2022}, which is not well understood yet. Furthermore, it is unclear how cause and effect work: is the saturated magnetic flux $\phi_{b,0}$ setting the magnetospheric radius $r_M$, or the other way around?

What happens outside the magnetospheric radius, $r>r_M$? If the monopolar magnetic field geometry continues, the magnetic pressure $p_b$ would become negligible compared to the gas pressure $p_g$ at $r>r_M$. However, accretion continues to bring in frozen-in magnetic fields which increase the importance of magnetic fields. We believe this continuous injection of magnetic fields is responsible for establishing equipartition, $p_b\approx p_g$, outside $r_M$. The required magnetic field scaling to balance gas pressure $p_g\propto r^{-2}$ is $B^r\propto r^{-1}$, which results in the magnetic flux scaling $\phi_b(r)\propto r^1$ outside $r>r_M$, with normalization given by
\begin{equation}
\phi_b(r) = \sqrt{8\pi^2 T_0} \left( \frac{r}{r_g} \right) \approx 2.8 \left( \frac{r}{r_g} \right), \qquad r>r_M.
\end{equation}
Thus, departing away from the monopolar field geometry is inevitable in order to accommodate accumulating magnetic fields and to achieve equipartition.

\subsection{Detailed Analysis of Large Amplitude Variability}\label{sec:eta_variation}

\begin{figure}[ht!]
\gridline{
\fig{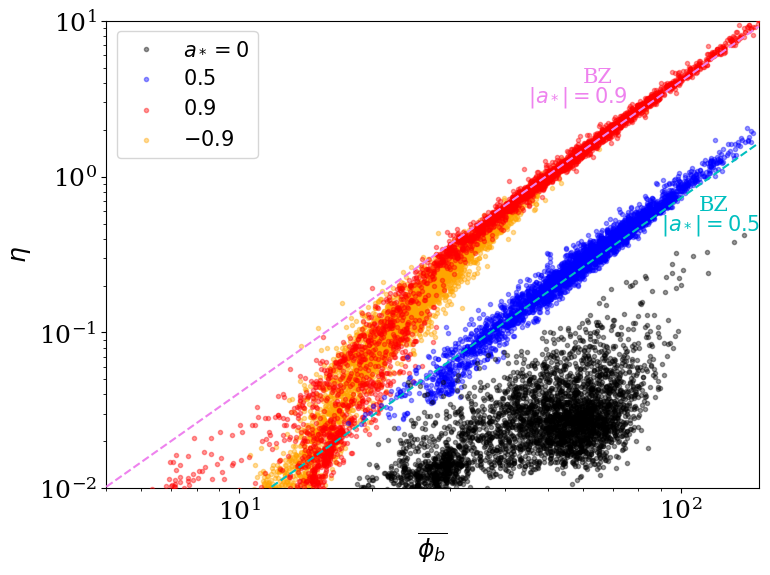}{0.48\textwidth}{}
}  
\caption{Same as \autoref{fig:standard eta vs phi} but for strongly magnetized torus $\beta_{\rm max}=1$. Despite the increased variability, the BZ $\eta-\phi_b$ relation (\autoref{eq:BZ}) is closely followed for magnetic flux $\phi_b\gtrsim 20$.
}\label{fig:strongly magnetized eta vs phi}
\end{figure}

In Section~\ref{sec:beta1_evolution} and \autoref{fig:strongly magnetized time evolution}, we noted the large variability we find in models which start with a strong initial magnetic field. Here we analyze the results in further detail.

We first check in \autoref{fig:strongly magnetized eta vs phi} the correlation between the efficiency parameter $\eta(10\,r_g)$ and the magnetization parameter $\phi_b(r_H)$. Even though the strongly magnetized tori considered here have a widely varying feedback efficiency, the BZ $\eta$-$\phi_b$ relation (\autoref{eq:BZ}) continues to hold well. The increased variability can be best seen for the $a_*=0.9$ run. In contrast to the standard $a_*=0.9$ run ($a.9\beta100$) where a high magnetic flux $\phi_b$ is maintained at all times (see the limited range of $\phi_b$ in the red dots in \autoref{fig:standard eta vs phi}), the highly magnetized counterpart here (run $a.9\beta1$, red dots in \autoref{fig:strongly magnetized eta vs phi}) goes through several periods of low magnetic flux, with $\phi_b$ falling even below $\sim10$. This large variation in $\phi_b$ leads to fluctuations in the feedback efficiency $\eta$ over 3 orders of magnitude. Even with such chaotic evolution, we confirm that the feedback for spinning BHs is governed by the BZ mechanism from the excellent agreement with \autoref{eq:BZ}. However, both \autoref{fig:standard eta vs phi} and \autoref{fig:strongly magnetized eta vs phi} show a similar deviation from the BZ relation when the magnetic flux is low, $\phi_b\lesssim 20$. 

\begin{figure}[]
\gridline{
\fig{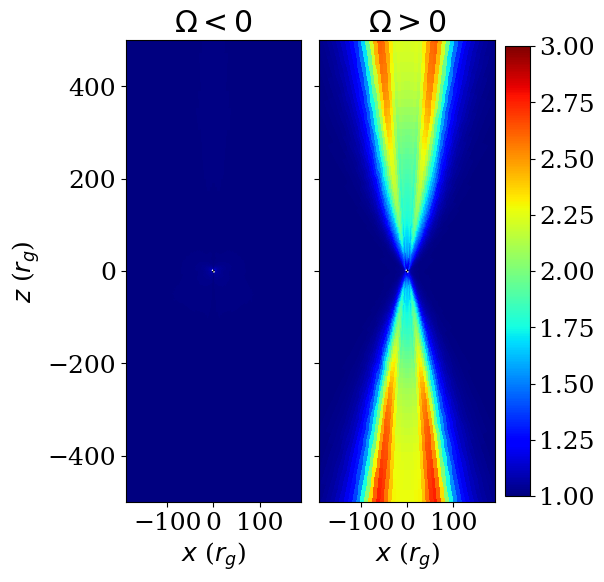}{0.4\textwidth}{}}
\caption{The distribution of the time- and $\phi$-averaged Lorentz factor $\Gamma$ in the poloidal plane in the strongly magnetized $a_*=0.9$ model ($a.9\beta1$) during two selected time periods: (\textit{left}) $t=1.3\times 10^5-1.6\times 10^5\,t_g$,  when fluid counter-rotates with respect to the BH spin and feedback efficiency is low, (\textit{right}) $t=2.5\times 10^5-2.8\times 10^5\,t_g$, when gas co-rotates with the BH and the feedback efficiency is high.}\label{fig:strongly magnetized jet gamma}
\end{figure}

There is a hint in the time evolution data in \autoref{fig:strongly magnetized time evolution} that flips in the sign of the angular velocity $\Omega$ may be relevant for understanding the large fluctuations in $\eta$ and $\phi_b$.
Especially for the prograde $a_*=0.9$ model (run $a.9\beta1$), during the period of negative angular velocity ($t=130000-160000\,t_g$), the accretion rate $\dot{M}$ is higher and the feedback efficiency is extremely small, $\eta\sim 3\,\%$, comparable to the efficiency of a nonspinning BH. This is confirmed in the time-averaged jet momentum slice in \autoref{fig:strongly magnetized jet gamma}, where during the same period of negative $\Omega$, the time-averaged $\Gamma$ in the left panel is non-relativistic, indicating that there is no jet activity. On the other hand, during periods of positive $\Omega$, e.g., $t=250000-280000\,t_g$, the efficiency is large and the jet is strong, as shown by the $\Gamma$ distribution in the right panel in \autoref{fig:strongly magnetized jet gamma}.

\begin{figure}[]
\gridline{
\fig{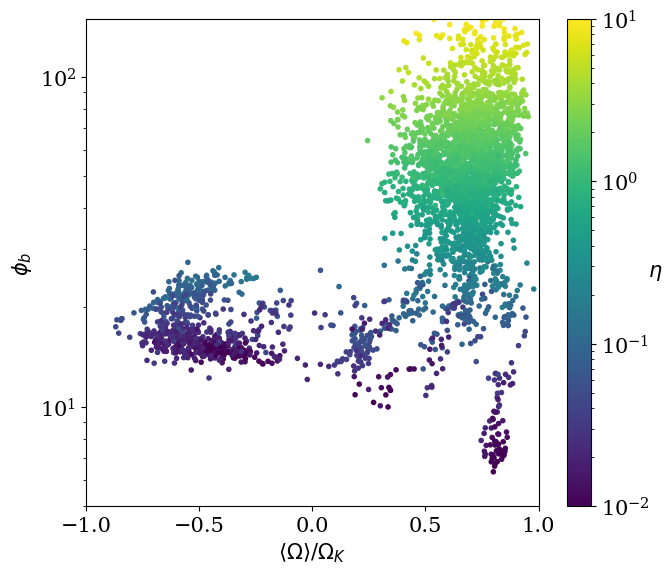}{0.45\textwidth}{}
            }   
\caption{Correlation plot between $\phi_b(r_H)$  and the shell-averaged fluid angular velocity $\langle\Omega\rangle/\Omega_K$ measured at $10\,r_g$ for the strongly magnetized model with spin $a_*=0.9$ (run $a.9\beta1$). The dots are colored by the efficiency $\eta(10\,r_g)$.
}\label{fig:strongly magnetized phiBH vs Omega}
\end{figure}

A correlation plot between $\phi_b(r_H)$ and the angular velocity $\langle\Omega\rangle/\Omega_K$ at $10\,r_g$ is shown in \autoref{fig:strongly magnetized phiBH vs Omega}, colored by the efficiency $\eta$ at $10\,r_g$. There is an interesting relation between the angular velocity $\Omega$ and the jet parameters $\eta, ~\phi_b$. In the case of weakly or counter-rotating gas ($\langle\Omega\rangle/\Omega_K \lesssim 0.2$), the magnetic flux and the efficiency are both low: $\phi_b<25$, $\eta\lesssim 0.1$. Similarly, the high feedback efficiency phase ($\eta \gtrsim 1$) always occurs when the gas is strongly co-rotating, $\Omega > 0.4\,\Omega_K$ (as similarly noted by \citealt{Galishnikova2025}). However, co-rotation of the gas does not always guarantee a high magnetic flux $\phi_b$ or efficiency $\eta$; we see in \autoref{fig:strongly magnetized phiBH vs Omega} that the feedback efficiency $\eta$ spans over 3 orders of magnitude from $\eta\sim 10^{-2}$ to $\sim 10^1$ for a positive $\Omega/\Omega_K\sim 0.8$.

\begin{figure}[ht!]
\gridline{
\fig{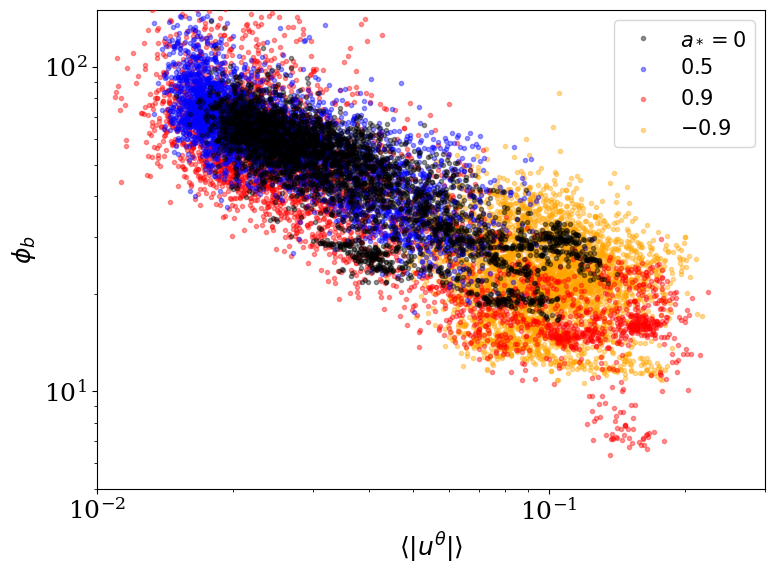}{0.45\textwidth}{}
            }   
\caption{Correlation between magnetic flux $\phi_b(r_H)$ and the polar velocity $\langle|u^\theta|\rangle(2\,r_g)$. There seems to be a cleaner correlation for all 4 spins (shown in different colors) compared to the correlation between $\phi_b$ and $\Omega$ in \autoref{fig:strongly magnetized phiBH vs Omega}.
}\label{fig:scatter_phib_uth}
\end{figure}

Interestingly, there is a stronger correlation between the magnetic flux $\phi_b(r_H)$ and the absolute value of the $\theta-$velocity $\langle |u^\theta|\rangle(2\,r_g)$ as shown in \autoref{fig:scatter_phib_uth}.\footnote{This correlation is also present when the velocity $\langle|u^\theta|\rangle$ is measured at a larger radius like $5\,r_g$ or $10\,r_g$.} The relation seems to be present for all BH spins; in fact, the points in \autoref{fig:scatter_phib_uth} join smoothly together and look like a single universal correlation, with $\phi_b$ decreasing as $|u^\theta|$ increases. We propose a possible explanation for the correlation. In the event of a strong flux eruption, the horizon magnetic flux $\phi_b(r_H)$ and the jet power drop temporarily. Then the jet cavity region, which was previously dominated by the magnetic pressure $p_b$, loses its support and the surrounding gas with a higher gas pressure $p_g$ floods into the cavity. This effect will show up as increased $\theta-$velocity $|u^\theta|$. 

The reason why a similar large variability in $\phi_b(r_H)$ is not observed for the weakly magnetized standard runs requires further investigation. It could be that for the strongly magnetized case, the lack of coherent rotation from magnetic braking paves the way for the flux eruptions to escape to larger distances and to carry away more horizon magnetic flux with them. On the other hand, for the weakly magnetized case, the coherent rotation suppresses outflow of flux eruptions and maintains a larger magnetic flux level near the horizon even immediately after a flux eruption. Another possibility is that the flux eruptions are just more powerful for the strongly magnetized torus.

\section{Comparison of Jet Intermittency and Large Fluctuations in Different Simulations}
\label{sec:comparison}

\begin{table*}[]
  \begin{center}
    \caption{Compilation of BH accretion simulations, along with their initial conditions (ICs),  characteristic radii $r_{\rm char}$, initial rotations, and initial plasma-$\beta$. The first row summarizes typical FM torus ICs used in most GRMHD studies. The FM torus setup used in \citet{Chan2025} and our present work are listed in the next two rows, with our choice of atypical torus size $r_{\rm char}$ and magnetic field strength $\beta$ highlighted in pink. The remaining studies adopt spherically symmetric Bondi-like ICs and are listed in succeeding rows. The initial rotational kinetic energy to magnetic energy ratio $\R$ is listed in the third to last column. An entry `O' in the second to last column (variability) indicates that at least one of the simulations in the particular study shows strong fluctuations in the magnetic flux $\phi_b$ (and jet efficiency $\eta$ if the BH is spinning). An entry `O' in the last column ($\Omega$ reversal) means that large changes in the gas angular velocity, including sign reversals, are observed; `X' indicates the effect is not detected, and blank means it is not reported.}
    \label{tab:comparison_prev_work}
    \begin{tabular}{c|c | c c c | c c c} 
    \hline
      Reference & IC & $r_{\rm char}/r_g$ & initial rotation & initial $\beta$ & $\R$ & variability & $\Omega$ reversal \\
      \hline
      \hline
      (Most GRMHD) &  & $\approx 20-40$  & $u^\varphi = u^\varphi_{\rm FM}$ & $100$ & $\gtrsim 500$ & X & X \\
        \citet{Chan2025} & FM torus  & $\approx 20$ & $u^\varphi=0-u^\varphi_{\rm FM}$& $\approx 1000$ & $0-2500$ & O &  \\
        \textbf{This work} &  & \colorbox{red!10}{$500$} & $u^\varphi=u^\varphi_{\rm FM}$ & $\colorbox{red!10}{1},100$ & $3, 300$& O & O \\
      \hline
        \citet{Ressler2021} &  & $100$ & 0 & 100 & 0 & O &  \\
        \citet{Kwan2023} &   & (not provided) & $r_c=0,10,50\,r_g$ & $>100$ & & O &  \\
        \citet{Lalakos2024} & & $10^3$ & 0 & 100 & 0 & O & \\
        \citet{Cho2024} & Bondi-like & $100-10^7$ & $u^\varphi = 0,0.9 \,r^{-3/2}$ & 1 & $0-0.5$ & O & O \\
        \citet{Galishnikova2025} &  & $50-2500$ & $r_c=2-50\,r_g$ & $10-10^4$ & $0-15$ & O & O\\
        \citet{Kim2025} & & $100$ & 0 & 10,100 & 0 & O & \\
        \citet{Guo2025} &  &  $250-2\times 10^5$ &  0 & 1,1000  & 0 & O & \\
        \citet{Lalakos2025} &  & $100-10^4$ & $r_c=0-300\,r_g$ & $\geq 100$ & $0-25$ & O & \\
        \hline
      \hline
      \end{tabular}
  \end{center}
\end{table*}

Our simulations with large-scale tori ($r_{\rm max}=500r_g$) give contrasting results on jet intermittency and variability amplitude. When the tori are initialized with a standard weak magnetic field ($\beta_{\rm max}=100$), we find steady strong jets with no sign of intermittency. However, when the tori are initially strongly magnetized ($\beta_{\rm max}=1$), the jet activity becomes intermittent and there is strong variability. Similar mixed outcomes have been reported in other recent studies, prompting us to systematically compare the different works and identify patterns.

\autoref{tab:comparison_prev_work} collects the available information on simulation setups and variability outcomes across multiple studies, including the present one. Some of the simulations are initialized with the FM torus (this includes most previous GRMHD simulations, which are combined into a single entry in the first row, and also the paper by \citet{Chan2025} and the present work). The remaining studies use spherically symmetric Bondi-like ICs.

In \autoref{tab:comparison_prev_work}, for the characteristic radius $r_{\rm char}$, we choose the radius of the pressure maximum $r_{\rm max}$ for simulations that use FM torus ICs, and the Bondi radius $r_B\equiv G M_\bullet /c_{s,\infty}^2$ for Bondi-like ICs, where $c_{s,\infty}=\sqrt{\gamma_{\rm ad}T_\infty}$ is the asymptotic sound speed at infinity. For the initial rotation, standard FM tori set this equal to the Keplerian angular velocity at the pressure maximum. Among the Bondi-like simulations, some consider only non-rotating gas ($u^\varphi=0$), while others select the initial rotation either by directly setting $u^\varphi$ or by parameterizing the initial rotation via the circularization radius $r_c$, in which case the azimuthal velocity at the midplane is initialized as $u^\varphi(r,\pi/2)= \sqrt{r_c}/r^2$. The column titled `initial $\beta$' includes the value of $\beta_{\rm max}$ for FM torus runs, and the initial $\beta$ at $r_B$ for the Bondi-like runs.

Note that the Bondi-like simulations vary widely in all three basic quantities: $r_{\rm char}$, initial rotation, initial $\beta$. The FM torus simulations also vary significantly, especially when the present work is included in the comparison.
As highlighted in pink background in \autoref{tab:comparison_prev_work}, our torus setup (third row) is different from the conventional FM torus (first row) setup in the characteristic size ($r_{\rm max}$) and the initial magnetic field strength ($\beta_{\rm max}$).
Comparing simulations with such diversity is challenging, so we introduce diagnostic parameters to summarize each simulation.

Our torus simulations show that rotation and magnetic fields affect each other in a nontrivial manner. Coherent rotation can slow down the outward motion of flux bundles following flux eruptions and thereby helps maintain a high level of horizon magnetic flux $\phi_b(r_H)$. On the other hand, strong magnetic fields weaken rotation, allowing stronger flux eruptions, which then overcome rotational shear further. Since there appears to be a competition between gas rotation and magnetic fields, we compare the magnetic energy density,
\begin{equation*}
    \epsilon_{\rm mag}(r,\theta) \equiv b^2/2 = p_g/\beta,
\end{equation*}
with the rotational kinetic energy density,
\begin{equation*}
    \epsilon_{\rm rot}(r,\theta) \equiv \rho r^2\Omega^2/2.
\end{equation*}
We define the ratio of these two quantities in the initial state at $r=r_{\rm char}$, $\theta=\pi/2$ as:
\begin{equation}
    \R\equiv \frac{\epsilon_{\rm rot}}{\epsilon_{\rm mag}}(r_{\rm char},\pi/2) = \left( \frac{\beta r^2\Omega^2}{2 T}\right)(r_{\rm char},\pi/2).
\end{equation}
The parameter $\R$ characterizes the relative importance of rotation compared to magnetic fields in the initial setup. Actually, accurately assessing buoyancy and convective instability, which drive flux bundles outward after flux eruptions, involves calculating radial derivatives of angular velocity and pressure (e.g., H\o iland criterion, \citealt{Balbus1995, Balbus2002}), but for simplicity we use $\R$ as an order-of-magnitude estimate.

For the FM torus at $(r_{\rm max},\,\pi/2)$, the angular velocity is nearly Keplerian, $\Omega \approx \Omega_K$, and the plasma-$\beta$ is $\approx \beta_{\rm max}$. Therefore, 
\begin{equation}
    \R = \frac{\beta_{\rm max}}{2T(r_{\rm max},\pi/2)\, r_{\rm max}}, \qquad (\text{FM torus}).\footnote{The temperature $T(r_{\rm max},\pi/2)$ is $\approx 3\times 10^{-4}$ in the present work,  and $\approx 0.01$ for \citet{Chan2025}. Also, \citet{Chan2025} artificially modify the rotation by a factor $f$ in order to explore its effect on jet variability, which requires an extra factor of $f^2$ in the equation.}
\end{equation}
For simulations with Bondi-like ICs, where the characteristic radius is the Bondi radius $r_B$, the asymptotic temperature is $T\approx (\gamma_{\rm ad}\, r_B)^{-1}$. The ratio $\R$ is then
\begin{align}
    \R &= \beta\gamma_{\rm ad} r_B^3\Omega(r_B,\pi/2)^2/2, \\
    &=\beta \gamma_{\rm ad} \frac{r_c}{2 r_B} ~ ~ (\text{Bondi-like}). 
\end{align}
In the last equality, we have rewritten the angular velocity in terms of the circularization radius $r_c$: $\Omega(r_B,\pi/2) = \sqrt{r_c}/r_B^2$. The third to last column in \autoref{tab:comparison_prev_work} gives our estimate of $\R$ for all the simulations considered here. 

The last two columns in \autoref{tab:comparison_prev_work} summarize the results from each work regarding variability in quantities like $\dot{M}$, $\eta$, $\phi_b$, $\langle\Omega\rangle$. The symbol `O' in the `variability' column indicates that at least one of the simulations in the paper showed large variability in $\dot{M}$, $\phi_b$, and $\eta$ (the efficiency is not relevant if the BH is not spinning, but variability in $\phi_b$ can still be used as a diagnostic of strong variability, e.g., the $a_*=0$ model in \autoref{fig:strongly magnetized time evolution}). The symbol `O' in the `$\Omega$ reversal' column indicates that there is large variability in $\langle\Omega\rangle$, including especially flips in the direction of gas rotation. The entry is left blank if the effect is not reported (we are not sure if the relevant authors looked for spin flips).

The conventional FM torus which is widely studied in the GRMHD literature (first line in \autoref{tab:comparison_prev_work}) has a large ratio $\R\gtrsim 500$, which points to the sub-dominance of magnetic fields compared to gas rotation in the initial state. The resulting evolution is relatively calm, accompanied (when BH spin is non-zero) by a persistent jet with little variability. On the other hand, simulations that are identified in \autoref{tab:comparison_prev_work} as having large variability or rotation flips are either initialized with zero rotation, hence $\R=0$, or have weak rotation and/or a strong initial magnetic field, hence a small value of $\R$.

\begin{figure}[ht!]
\gridline{
\fig{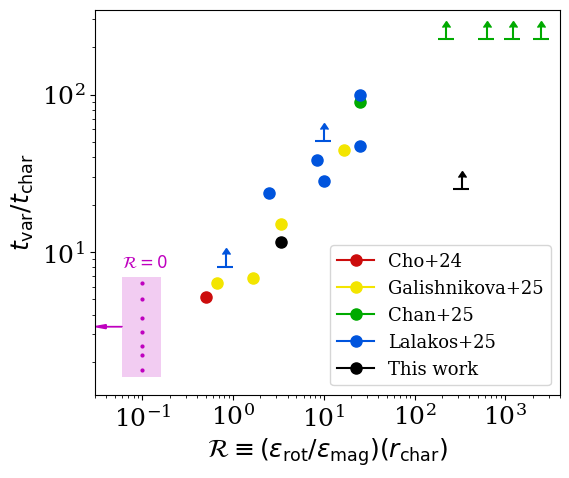}{0.45\textwidth}{}
}
\caption{Plot of the normalized time at which strong variability (in $\phi_b$, $\eta$ or $\Omega$) is first observed, $t_{\rm var}/t_{\rm char}$, against the initial rotational to magnetic energy ratio, $\R$, for the studies included in \autoref{tab:comparison_prev_work}. Simulations which do not show any strong variability until the end of the run are indicated as $t_{\rm var}$ bounded below by the total runtime. Simulations that are initialized with non-zero rotation are shown as larger symbols, color-coded by the relevant reference. Simulations that start with no rotation, $\R=0$, are shown as small pink dots in the left corner. The data show a clear correlation: Simulations that begin with a larger initial rotational dominance over magnetic energy take a longer time before they show strong variability.}
\label{fig:compare_tvariable}
\end{figure}

\citet{Cho2024}'s run with a small $\R\approx 0.5$ ($r_B\approx 400\,r_g, ~\beta\sim 1, ~\Omega\approx u^\varphi =0.9\,r^{-3/2}$) found both large variability in $\phi_b$ and angular velocity flipping. \citet{Galishnikova2025}'s runs with $r_B=250\,r_g$, $\beta=100$, and various values of $r_c$ have $\R$ ranging from 0 to $\sim 15$. They report that all their simulations eventually reach intermittent jet behavior. Interestingly, models with faster initial gas rotation (larger $r_c$, hence large $\R$) take a longer time to achieve strong variability. 
\citet{Chan2025}'s FM torus simulations ($r_{\rm max}\approx 20, ~\beta_{\rm max}\approx 1000$), with angular velocity $u^\varphi$ reduced relative to the equilibrium FM torus, show large variability when the initial angular velocity is $0.1$ times the FM torus value $u^\varphi_{\rm FM}$ (corresponding to $\R\approx 25$) and even more robust variability for $u^\varphi = 0$ ($\R=0$). However, for more rapidly rotating models  with $u^\varphi\geq 0.3\,u^\varphi_{\rm FM}$, the \citet{Chan2025} tori behave like conventional FM tori with a steady jet (no intermittency). Likewise, \citet{Lalakos2025}'s models with $\R=0-25$, which use spherically symmetric ICs with $r_B=1000\,r_g$ and initial $\beta\geq 100$, take a longer time before exhibiting strong jet variability as $r_c$ (hence $\R$) is increased. In the same vein, our weakly magnetized `standard' torus runs (Section~\ref{sec:standard models}), which are described by a large value of $\R\approx 300$, do not show strong variability until the end of the simulation ($t=2.8\times10^5t_g$, see \autoref{fig:standard time evolution}), while our strongly magnetized torus runs (Section~\ref{sec:beta1_models}), which have a small $\R\approx 3$, exhibit large amplitude variability after $t\sim10^5t_g$ (\autoref{fig:strongly magnetized time evolution}). 

Motivated by the above discussion, we introduce a second key parameter: the time $t_{\rm var}$ at which a given simulation first exhibits strong variability. To make this quantity dimensionless, we scale it by the characteristic time, which we take to be the free-fall time at the characteristic radius, $t_{\rm char}\equiv (r_{\rm char}/r_g)^{3/2}\,t_g$. For those simulations that do not show strong variability, e.g., our standard torus runs in Section~\ref{sec:standard models}, we take the total duration of the simulation to be a lower bound on $t_{\rm var}$.

\autoref{fig:compare_tvariable} is a plot of $t_{\rm var}/t_{\rm char}$ against $\R$ for all the simulations in \autoref{tab:comparison_prev_work} for which both parameters can be extracted from the published information.
In the case of \citet{Cho2024} and \citet{Guo2025}, for a fair comparison, we only use their conventional GRMHD simulations (referred to as one-zone or ground truth) and ignore their accelerated simulations which make use of the multi-zone or cyclic zoom schemes, where the simulation time is less easy to interpret. In addition, we leave out some of \citet{Ressler2021}'s simulations where the initial magnetic field is misaligned with the BH spin axis, and also \citet{Kim2025}'s Bondi-Hoyle simulations where the BH travels through a uniform fluid.

Among the remaining simulations, those which begin with no initial rotation ($\epsilon_{\rm rot}=0$) are problematic because all information about the magnetic energy $\epsilon_{\rm mag}$ is lost when calculating $\R$. Presumably, the evolution of these systems is influenced by some (unknown) factor other than rotation whose importance relative to $\epsilon_{\rm mag}$ may play a role. Because of this uncertainty, we show these data in \autoref{fig:compare_tvariable} as small pink dots. Additionally, \citet{Galishnikova2025} report that variability is unphysically suppressed in their Bondi-like simulations if the simulated system is too small in size, e.g., $r_B = 50\,r_g$. The small volume apparently applies strong geometrical constraints. We thus restrict our attention to larger systems with $r_{\rm char} \geq 100\,r_g$ for $\R=0$ runs.

In the case of initially rotating models, $\R>0$ is a meaningful parameter, and we plot the corresponding data in \autoref{fig:compare_tvariable} with large symbols, using different colors for each study.  In those cases where a simulation found no large amplitude variability until the end of the run, we show the runtime divided by  $t_{\rm char}$ as a lower bound on the variability time. In view of \citet{Galishnikova2025}'s result that very small systems may be unnaturally steady, one might wish to downweight the results from \citet{Chan2025} whose initial tori have $r_{\rm max}\approx 20\,r_g$. But their results agree with the other studies, so we include them for completeness.

Despite the fact that the various simulations under consideration vary widely in their setups, and that $\R$ is only a crude estimate of the initial importance of rotation relative to magnetic fields, the data plotted in \autoref{fig:compare_tvariable} show a clear trend of increasing $t_{\rm var}/t_{\rm char}$ with increasing $\R$. The initially non-rotating $\R=0$ simulations shown in the bottom left corner of  \autoref{fig:compare_tvariable} all reach strong variability over a time $t_{\rm var}\sim 1- 7\, t_{\rm char}$. The small difference in $t_{\rm var}/t_{\rm char}$ in these studies likely originates from other details of the initial setup that $\R$ does not encapsulate.

Inferring from the trend in \autoref{fig:compare_tvariable}, we predict that even our weakly magnetized tori ($\R\approx 300$), which remain stable until the end of the simulations ($t=2.8\times10^5t_g$), will finally end up with strong variability and intermittent jets if the simulations are run more than $10$ times longer. Simulating for such extended times is computationally expensive but can be achieved with some newly developed numerical schemes \citep{Cho2023,Cho2024,Guo2025}.

Physically, the switch to a state with strong fluctuations and jet intermittency at $t_{\rm var}$ is likely due to the accumulation of enough magnetic flux at the center to the point of efficiently braking the rotation of the accreting gas and powering strong flux eruptions which have less rotation and shear to overcome.
For a rapidly rotating FM torus, we note that a large size (like the models in this paper) may be required to verify this picture; a small torus with a limited amount of gas and magnetic fields might go through severe gas depletion before it has the opportunity to switch states.

If our prediction is correct that all the systems we have considered here will ultimately become violently variable and develop intermittent jets, then it implies that the majority of real systems in Nature with hot accretion flows must also behave in a similar fashion. This is because even the longest simulations we can run are far shorter than the lifetimes of real systems.

\section{Summary and Conclusion}\label{sec:conclude}
In this paper, we investigated the strong fluctuations in jet power and the reversals in angular velocity recently observed in some GRMHD simulations of BH accretion---a feature  largely absent in previous simulations initialized with the FM torus.
Most of the newer studies finding strong variability \citep{Ressler2021, Kwan2023, Lalakos2024, Cho2024, Galishnikova2025, Kim2025, Guo2025,Chan2025,Lalakos2025} use spherically-symmetric, Bondi-like ICs, and differ from typical FM torus setups in several aspects, including characteristic size, initial rotation, and magnetic field strength. To isolate the key drivers of variability, we ran simulations in which we modified the FM torus ICs in two different respects.

First, in Section~\ref{sec:standard models}, we increased the size of the torus by more than an order of magnitude, setting the pressure maximum at $r_{\rm max}=500\,r_g$, compared to the typical $r_{\rm max}\approx 20-40\,r_g$ in most previous torus simulations. We retained the weak initial magnetic fields ($\beta_{\rm max}=100$) traditionally used in such work. The larger torus size allows long-term evolution without gas depletion, avoiding a problem with small FM tori. However, other than this, the large torus runs reproduce small fluctuations of the previous small tori simulations, thereby validating the earlier results. In particular, based on the strong correlation between the jet efficiency $\eta$ and the magnetic flux parameter $\phi_b$ (\autoref{fig:standard eta vs phi}, also \autoref{fig:strongly magnetized eta vs phi} in Section~\ref{sec:beta1_models}), we confirm that jet power is driven by the BZ mechanism. 

Two new results from these large torus simulations are the following: (i) We observe a moderate spin dependence of the accretion rate $\dot{M}$ on the BH, with reductions by factors of $\sim1.6$ and 2.5 for black hole spins $a_*=0.5$ and 0.9, respectively, relative to the non-spinning case (this is confirmed also in Section~\ref{sec:beta1_models}, \autoref{fig:Mdot_comparison}). (ii) The extended radial range of the larger torus allows an improved measurement of the density scaling: $\rho(r) \propto r^{-1.1}$ for prograde spins $a_*\geq 0$, and $\rho(r) \propto r^{-0.8}$ for the retrograde $a_*=-0.9$ model (\autoref{fig:density_comparison}). 

In the simulations in Section~\ref{sec:standard models}, the evolution of various quantities, including jet efficiency $\eta$, shows only small fluctuations around the time-averaged values, and the angular velocity $\Omega$ of the accreting gas maintains a consistent sign (\autoref{fig:standard time evolution}). We thus verify that having a larger dynamic range of radius via a larger initial FM torus does not by itself produce the large variability found in other studies. 

In Section~\ref{sec:beta1_models}, we introduce a second major change to the FM torus setup---we initialize the torus with a strong magnetic field ($\beta_{\rm max}= 1$) instead of the standard weak field ($\beta_{\rm max}= 100$).  Accretion begins earlier in these models due to more efficient angular momentum transport, and all the models show significant variability across BH spins. Strong flux eruptions produce low-density bubbles extending to $r\sim 3000\,r_g$, an order of magnitude larger than the $r\sim 200\,r_g$ seen in weakly magnetized runs. 

We analyze the shape of $\phi_b(r)$. We find that it is flat, indicative of a split monopole geometry, up to a magnetospheric radius $r_M \sim 10\,r_g$, and we derive an analytical formula (\autoref{eq:rm_phib}) connecting the horizon flux $\phi_b(r_H)$ to $r_M$. Beyond $r_M$, we find $\phi_b(r) \propto r$, suggesting a change in the field geometry. This zone, which extends up to  $r \sim 3000\,r_g$, corresponds to a radially extended magnetically arrested disk (MAD) state.

In the strongly magnetized torus runs, we find intermittent jet behavior correlated with angular velocity reversals, particularly in the $a_*=0.9$ model. When the jet is strong ($\eta, ~\phi_b$ large), the gas rotates rapidly, with $\Omega > 0.4\Omega_K$; when the jet is weak, the rotation can be anywhere between strongly co- and counter-rotating (\autoref{fig:strongly magnetized phiBH vs Omega}). Comparing with previous Bondi-like simulations, we note that angular velocity reversals were reported only in \citet{Cho2024} and \citet{Galishnikova2025}, but it may have been present in other studies as well, since it is unclear if the authors studied the evolution of $\Omega$. We also identify a clear correlation between the polar velocity $|u^\theta |$ and the magnetic flux parameter $\phi_b$, where $|u^\theta |$ increases as $\phi_b$ decreases (\autoref{fig:scatter_phib_uth}). We interpret this as follows: during strong flux eruptions, magnetic pressure drops, allowing surrounding gas (with higher thermal pressure) to flood the jet cavity, increasing $|u^\theta|$ and further suppressing the jet.

To facilitate comparisons among diverse simulations and understand better the conditions under which strong jet variability and rotation flips occur, we introduce two diagnostic parameters:
\begin{itemize}
    \item $\R$: the ratio of rotational kinetic to magnetic energy in the initial state,
    \item $t_{\rm var}/t_{\rm char}$: the onset time of strong variability normalized by the characteristic timescale.
\end{itemize}
Plotting $t_{\rm var}/t_{\rm char}$ against $\R$ across all initially rotating simulations reveals a consistent trend (\autoref{fig:compare_tvariable}): larger $\R$ delays the onset of variability. This delay likely reflects the time required for magnetic fields to build up sufficiently to overcome rotational shear and drive significant flux eruptions. Our strongly magnetized torus ($\R\approx 3$) transitions to strong variability at $t_{\rm var}/t_{\rm char} \sim 10$, consistent with Bondi-like runs with similar values of $\R$. Our weakly magnetized torus ($\R\approx 300$) maintains a steady jet through the end of the run, but based on the trend, we predict that extending the simulation by more than 10 times longer would eventually trigger variability.

We conclude by highlighting our key results:
\begin{itemize}

\item Increasing the size of the initial FM torus enables long-term evolution without gas depletion and improves measurements of radial structure---we find density scaling as $\rho(r)\propto r^{-1.1}$ for prograde BH spins. However, a larger torus alone does not produce strong jet variability.

\item 
Introducing strong magnetization in the initial torus ($\beta_{\rm max}=1$) leads to large-amplitude fluctuations in $\eta$ and $\phi_b$, and thus intermittent jets. It also produces angular velocity reversals, which coincide with periods of low $\eta$ and $\phi_b$.

\item
For similar boundary conditions, the mass accretion rate $\dot{M}$ on the BH decreases modestly with increasing BH spin.

\item
We provide a physical interpretation of the radial behavior of $\phi_b(r)$, and derive a relation linking the horizon flux $\phi_b(r_H)$ to the magnetospheric radius $r_M$.

\item A clear trend across studies shows that larger values of the parameter $\R$ delays strong variability. This supports the idea that jets in all magnetized hot accretion flows eventually become intermittent, but the required time depends on the initial energy balance between rotation and magnetic fields. Since BH jets in hot accretion systems in Nature live far longer than the physical time covered in any simulation, they should all be intermittent.

\end{itemize}

By modifying the FM torus in a controlled way, in this paper we have reproduced the variability seen in simulations with Bondi-like initial conditions. Our results offer a clean framework for disentangling the mechanisms behind jet intermittency and open the door for further investigation under more simplified and controlled conditions.

\section*{Acknowledgements}
HC and RN were partially supported by the Black Hole Initiative at Harvard University, which is funded by the Gordon and Betty Moore Foundation (Grant \#8273.01). It was also made possible through the support of a grant from the John Templeton Foundation (Grant \#62286).  The opinions expressed in this publication are those of the authors and do not necessarily reflect the views of these Foundations.
This work used Anvil at Purdue University through allocation AST080028 from the Advanced Cyberinfrastructure Coordination Ecosystem: Services \& Support (ACCESS) program, which is supported by U.S. National Science Foundation grants \#2138259, \#2138286, \#2138307, \#2137603, and \#2138296.

\clearpage
\newpage

\bibliography{bigtorus}{}
\bibliographystyle{aasjournal}

\end{CJK}
\end{document}